\documentclass[%
 aip,
 amsmath,amssymb,
 reprint,%
]{revtex4-1}

\usepackage{graphicx}% Include figure files
\usepackage{dcolumn}% Align table columns on decimal point
\usepackage{bm}% bold math
\usepackage{xcolor}
\usepackage[export]{adjustbox}
\usepackage[utf8]{inputenc}
\usepackage[T1]{fontenc}
\usepackage{mathptmx}
\usepackage{float}
\usepackage{subfig}
\usepackage{multirow}
\usepackage{array}
\usepackage{etoolbox}
\usepackage{mathastext}
\usepackage{tabularx}
\usepackage{amsmath}
\makeatletter
\def\@email#1#2{%
 \endgroup
 \patchcmd{\titleblock@produce}
  {\frontmatter@RRAPformat}
  {\frontmatter@RRAPformat{\produce@RRAP{*#1\href{mailto:#2}{#2}}}\frontmatter@RRAPformat}
  {}{}
}%
\makeatother
\begin{document}

\preprint{AIP/123-QED}

%\title{State-to-state and reduced-order model analysis of recombination in non-equilibrium $N_{2}+N$ and $O_{2}+O$ 0-D simulations}
\title{Master equation study of three-body recombination of nitrogen and oxygen in non-equilibrium hypersonic flows}
% Force line breaks with \\

\author{Aakanksha Notey}
\altaffiliation[PhD student]{}
\affiliation{Center for Hypersonics and Entry Systems Studies (CHESS), \\ University of Illinois at Urbana-Champaign, Urbana, IL 61801, USA}

\author{Sung Min Jo}
\altaffiliation[Assistant Professor]{}
\affiliation{Department of Aerospace Engineering, \\ Korea Advanced Institute of Science and Technology (KAIST), Daejeon 34141, Republic of Korea}

\author{Marco Panesi}
\altaffiliation[Professor, Corresponding author]{}
\email{mpanesi@uci.edu}
\affiliation{Center for Hypersonics and Entry Systems Studies (CHESS), \\ University of Illinois at Urbana-Champaign, Urbana, IL 61801, USA}
\affiliation{Department of Mechanical and Aerospace Engineering, \\ University of California Irvine, Irvine, CA 92697, USA}

\date{\today}% It is always \today, today,

\begin{abstract}
This work aims to study the energy transfer and recombination processes in N$_{2}$$\left(^{1}\sum^{+}_{g}\right)$+N$\left(^{4}S_{u}\right)$ and O$_{2}$$\left(^{3}\sum^{+}_{g}\right)$+O$\left(^{3}P_{2}\right)$ chemical systems when the system is suddenly cooled in a 0-D isothermal reactor thereby inducing strong non-equilibrium. A state-to-state (StS) study of the non-equilibrium phenomenon is crucial for developing accurate and efficient reduced-order models that can accurately capture the thermophysics involved. The gas mixture, consisting primarily of atoms at a high initial temperature of 10,000 K, is suddenly plunged into a low-temperature heat bath to simulate non-equilibrium recombination conditions. The population distribution of microscopic energy levels for each system is determined by solving a system of master equations. The conventional assumption of faster equilibration of rotational mode as compared to the vibrational mode holds for $ N_2 +N$, while it is not a very strong assumption for $ O_2 +O$ as the two relaxation time scales are comparable. Effective recombination rate constants for the quasi-steady state (QSS) period are calculated using the population distribution obtained by solving the master equations. It was also observed that the relaxation time constants for heating and cooling are different, with the time constant being lower for the cooling case due to anharmonicity effects in expanding flows. An attempt has also been made to use the insights from the StS analysis to determine an accurate binning strategy for the recombination processes involved in the two chemical systems. 
\end{abstract}

\maketitle

\section{\label{sec:level1} INTRODUCTION}

Hypersonic flows exhibit strong shock waves and significant viscous interactions, resulting in high post-shock temperatures that are sufficient to drive chemical reactions. These flows are characterized by steep gradients and rapid transients, especially at high altitudes where low densities and short residence times hinder equilibration, making thermochemical nonequilibrium effects critical. While shock-induced compression primarily leads to dissociation, it is in expansion regions—such as off the shoulder of blunt bodies, in wakes, and within nozzles—that recombination dominates, and nonequilibrium chemistry plays an even more pivotal role \cite{Park}.

Recombination of dissociated species in these expanding flows leads to localized heat release, significantly influencing surface heat transfer, particularly at cold wall boundaries. Accurate modeling of this process is essential not only for designing thermal protection systems but also for interpreting diagnostics in high-enthalpy ground test facilities. In such facilities, low-density nozzle flows often freeze chemically, deviating from equilibrium, and impacting the validity of experimental results.

Furthermore, recombination phenomena affect multiple mission-relevant observables: chemiluminescence in wakes (e.g., NO radiation in meteor trails), electron densities relevant to radio blackout prediction, and modifications to aerodynamic and thermal loads in the wake regions of vehicles \cite{Park}. Understanding and modeling these effects are thus foundational to predictive simulation and experimental validation of hypersonic flight environments \cite{Gnoffo, Park}.

Existing nonequilibrium models span computationally efficient empirical formulations \cite{Adamovich, SSH_Approx, Zener, Parkmodel, Boyd_emp_sts, Kim_Hydrogen} to computationally intensive state-to-state (StS) models derived from ab-initio quantum chemistry data \cite{Panesi2013_N3, Sung_Min_N2O, Kim_Hydrogen, Boyd_Sts_N2, Boyd_O2, Boyd_emp_sts, Macdonald2020_N3}. Empirical models, such as the widely adopted Park Two-Temperature model \cite{Parkmodel}, often rely on simplifying assumptions that limit their accuracy across diverse conditions. This model uses a single temperature $T$ for translational-rotational modes and a second temperature $T_{ve}$ for vibrational-electronic modes, including translational energies of free electrons. Park's model struggles at high temperatures, particularly under translational-rotational nonequilibrium conditions. Conversely, accurate StS models are computationally prohibitive for routine CFD applications, necessitating reduced-order models (ROMs) that accurately describe the nonequilibrium effects with manageable computational costs.

Previous research has extensively explored reduced-order models for nonequilibrium hypersonic flows \cite{Venturi2020_O3Diss, Munafo2020_N2O, Sahai2017_Adaptive, Macdonald2018_CGQCT, Macdonald2018_CGQCT2, Robyn_N3_CGQCT, Colonna_TLD_2011, Colonna_TLD_2012,Zanardi_PG,Zanardi_Ar}. Typical ROM approaches, such as the multi-group maximum entropy model \cite{Liu2015}, group energy levels into bins based on equilibration likelihood. Although existing binning strategies (vibration-specific and energy-based) \cite{Macdonald2018_CGQCT2, Macdonald2018_CGQCT, Sung_Min_N2O, Venturi2020_O3Diss} do not require prior knowledge of dissociation or recombination phenomena, physics-informed ROM development that accurately reproduces StS model results relies heavily on insights from detailed StS studies. Prior research has primarily focused on nonequilibrium dissociation processes within kinetic databases for $N_{2}+N$ \cite{Panesi2013_N3,Boyd_Sts_N2} and $O_{2}+O$ \cite{Boyd_O2,Boyd_O21}. However, recombination phenomena have received comparatively less attention. In one of the previous studies on recombination, Munaf\`o et. al. \cite{Munafo_recomb} applied StS collisional and maximum-entropy based model-reduction to simulate recombination and energy transfer in an expanding nozzle flow of the Electric Arc Shock Tube Facility (EAST) at NASA Ames Research Center by solving master equations. The vibrational specific state-to-state method has also been previously applied to CFD solvers to study the recombination phenomenon in nozzle flow \cite{Capitelli_BL,Colonna_1999,Colonna_2001} and boundary layers \cite{Capitelli_BL}. A vibrational State-to-State approach was applied to DSMC collisional models by Pan et. al \cite{VibStS_Recomb_DSMC} to study the energy exchange dynamics in $\mathrm{O_{2}+O}$ in dissociating and recombining environments. Recently, Macdonald conducted an analysis of population distribution evolution and derived macroscopic quantities for $N_{2}+N$ and $O_{2}+O$ systems under recombination conditions \cite{Robyn_recomb} using rovibrational StS and vibrational specific models. This article concluded that the recombination is essentially vibration-specific as the vibrational distribution of the energy levels is not significantly affected upon omission of rotational non-equilibrium consideration. While this article discussed the macroscopic quantities like mole fraction, internal temperature and recombination rate constant evolution in detail, a comprehensive study of microscopic population distribution evolution and macroscopic relaxation times associated with the energy transfer process is missing. A knowledge of these quantities is of utmost importance while developing reduced-order models. To bridge this knowledge gap, this paper aims to perform a detailed StS analysis of recombination and energy transfer processes in $N_{2}+N$ and $O_{2}+O$ chemical systems to characterize their behavior in thermochemically nonequilibrium states and use the insights obtained to develop a novel binning strategy for model-reduction.

%\textcolor{red}{SAY THAT VENTURI WAS THE FIRSt to DO RECOMBINATION BEFORE MACDONALD EVEN, ALSO, I THINK MY WIFE HAS ALSO A PAPER on RECOMBINATION}\textcolor{red}{(ADD CAPITELLI GROUP REFERENCES HERE, THEY DID STUDY RECOMBINATION BUT USING VIB SPECIFIC MODEL)}

%\textcolor{blue}{What do you add to robyn's research, you need to write something about it.}

The paper structure is as follows: Section II describes potential energy surfaces employed in QCT calculations, thermodynamic and kinetic databases, the system of master equations, macroscopic parameters (e.g., recombination rate constants and internal energy), and reduced-order modeling methodologies. Section III presents the results, organized into four subsections: (A) analysis of dissociation/recombination processes; (B) study of energy transfer processes independent of dissociation/recombination; (C) evaluation of macroscopic recombination rate constants; and (D) assessment of reduced-order modeling outcomes employing various binning strategies.

\section{PHYSICAL MODELING}
This section outlines the potential energy surfaces, thermodynamic and kinetic databases, and the rovibrational collisional and reduced-order model equations used to analyze the recombination phenomenon in the two chemical systems. 

\subsection{Potential energy surfaces}
The rovibrational collisional model calculates the temporal evolution of the molecular population by solving a system of master equations. The microscopic dissociation/recombination and energy transfer rate constants used to solve the master equations are obtained by using the quasi-classical trajectory (QCT) method \cite{Capitelli_O3_QCT, Bender_QCT}. The rate coefficients are obtained from cross-sections by simulating numerous collisions between particles to determine the probability that a particular reaction will occur. The QCT method assumes that the dynamics of collision occur classically, while the initial and final states are obtained by mapping discrete quantum states. The QCT calculations are performed by CoarseAIR \cite{Venturi2020_O3Bayesian,Venturi2020_O3Diss}, an in-house FORTRAN code based on the original vectorized variable step size trajectory code (VVTC) developed by Schwenke \cite{Schwenke1988_VVTC}at NASA Ames Center. 
The QCT cross-sections require a potential energy surface (PES) for each chemical system as described in the subsections below. 

\subsubsection{$N_{2}+N$ PES}
In this work, the $N_{2}+N$ PES developed by NASA Ames Research Center \cite{Jaffe_N3_PES1,Chaban_N3_PES2} was used to carry out the QCT calculations. The kinetic databases obtained consider both the nitrogen molecule and the colliding nitrogen atom to be in their ground state. Therefore, the present work specifically analyzes the N$_{2}$$\left(^{1}\sum^{+}_{g}\right)$+N$\left(^{4}S_{u}\right)$ chemical system. The total rovibrational energy levels of the nitrogen molecule $N_{2}(v,J)$ , where $v$ stands for vibrational quantum number and $J$ stands for rotational quantum number, are 9399 with 61 vibrational levels ranging from 0-60 for the rotationless potential (J=0). The maximum rotational quantum number for the ground vibrational level (v=0) is 279. The database comprises 7428 bound levels, with the dissociation energy of the electronic ground-state molecular nitrogen being 9.75 eV. The remaining 1971 levels, known as the quasi-bound levels, have energy higher than the dissociation energy but lower than the centrifugal barrier energy value for the specific rotational quantum number. The symmetric wave function of nitrogen molecules does not change sign upon exchange, thereby exhibiting Bose-Einstein statistics with the degeneracy of energy levels given by $(2J+1)g^{NS}$ where $g^{NS}$ is the nuclear spin degeneracy \cite{Panesi2013_N3}. The nuclear spin degeneracy is 6 for even rotational states and 3 for odd rotational states. 

\subsubsection{$O_{2}+O$ PES}
The StS analysis of the $O_{2}+O$ chemical system uses the ground state O$_{2}$$\left(^{3}\sum^{+}_{g}\right)$+O$\left(^{3}P_{2}\right)$ PES developed by Varga et.al \cite{Varga_O3_PES}. The QCT method used is similar to that described above. The StS collision cross sections are calculated for a range of collision energies from 0.1 to 30 eV. The oxygen database comprises 6115 rovibrational energy levels, $O_{2}(v,J)$ with 45 vibrational states and a maximum rotational quantum number of 240. The dissociation energy of the electronic ground state of molecular oxygen is 5.113 eV with 4580 bound levels. The degeneracy of energy level is $g_{e}g_{i}$ with $g_{e}=3$ being the ground electronic state degeneracy and $g_{i}=\frac{1}{2}(2J+1)$ being the rovibrational degeneracy. The rovibrational degeneracy has a factor of $1/2$ to incorporate the nuclear spin contribution based on a semi-classical approximation, as the oxygen molecule has only odd rotational states \cite{Venturi2020_O3Diss}.

\subsection{Kinetic and thermodynamic databases}
The collisional processes considered for the two systems are: excitation/de-excitation and dissociation/recombination of $N_{2}$ and $O_{2}$ due to collisions with N and O, respectively.\\
The two processes are represented as follows:

Excitation and de-excitation : $A^{i}+B \leftrightarrow A^{j}+B $

Dissociation and recombination : $A^{i}+B \leftrightarrow C+D+B $
\\
where superscripts $i$ and $j$ refer to the rovibrational levels. The species A with a superscript denotes molecules, and the species B, C, D without a superscript indicate bulk atomic species without tracking their internal state's collisional transitions. The molecular species A is composed of atomic species C and D. The present work considers homogeneous chemical systems; therefore, in this study, $ B=C=D$, i.e., the atomic species are the same. The excitation/de-excitation reaction rate coefficients can be obtained for either exothermic or endothermic processes from QCT calculations. The exothermic rate constants were used in the present study for both chemical systems. The corresponding endothermic rate constants are obtained by imposing micro-reversibility on the exothermic rate constants as given below:
\begin{align}
    k_{ji}(T)=k_{ij}(T)\frac{g_{i}}{g_{j}} exp\left[\frac{\left(E_{j}-E_{i}\right)}{k_{B}T}\right],
\end{align}
where $k_{ij}$ and $k_{ji}$ refer to the forward and backward rate constants respectively, $g_{i}$ and $g_{j}$ are the degeneracies of the rovibrational levels, $k_{B}$ is the Boltzmann constant and T is the translational temperature. Similarly, using the principle of micro-reversibility, the recombination rate constants may be obtained from the dissociation rate constants:
\begin{align}
    k_{i}^{r}(T)=k_{i}^{d}(T)\frac{g_{e}g_{i}Q_{A}^{t}(T)}{\left[g_{B}Q_{B}^{t}(T)\right] ^{2}} exp\left(\frac{2E_{B}-E_{i}}{k_{B}T}\right),
\end{align}
where $g_{B}$ is the atom degeneracy and has values 12 and 9 for N and O respectively, $g_{e}$ is 1 and 3 for N and O respectively, and $E_{B}$ denotes the formation energy of the B atom. The translational partition function for the molecular and atomic species is defined as:
\begin{align}
    Q^{t}_{A,B}(T)=\left(\frac{2\pi k_{B}m_{A,B}T}{h_{P}^{2}}\right)^{3/2},
\end{align}
where $h_{P}$ stands for Planck's constant.

\subsection{Master equations}
A system of master equations comprising a species continuity equation for the nitrogen/oxygen atom and the rovibrational energy levels of the molecule is used to determine the evolution of the population distribution of molecules in a particular rovibrational energy level $i(v,J)$ \cite{Panesi2013_N3}:

\begin{align}
\begin{split}
    \frac{dn_{A}^{i}}{dt}= \sum_{j}\left(-k_{ij}n_{A}^{i}n_{B}+k_{ji}n_{A}^{j}n_{B}\right)+\left(-k_{i}^{d}n_{A}^{i}n_{B}+k_{i}^{r}n_{B}n_{C}n_{D}\right),
\end{split}
\end{align}

\begin{align}
    \frac{dn_{B}}{dt}=2\sum_{i}\left(k_{i}^{d}n_{A}^{i}n_{B}-k_{i}^{r}n_{B}n_{C}n_{D}\right),
\end{align}
where $n_{A}^{i}$ represents the number density/population of internal level i of species A and $n_{B}$ represents the number density of the atomic species. The total number density of species A is then obtained by summing over all its internal levels.

The population of rovibrational energy levels is initialized according to a Maxwell-Boltzmann distribution:
\begin{equation}
    \frac{n_{i}}{n_{A}}=\frac{g_{e}g_{i}}{Q^{I}(T^{I})}exp\left(\frac{-E_{i}}{k_{B}T^{I}}\right),
\end{equation}
at the initial internal temperature $T^{I}$. The molecular internal partition function is defined as $Q^{I}(T^{I})=\sum_{j}g_{e}g_{j}exp\left(-E_{j}/(k_{B}T^{I})\right)$ and the number density of molecule at any instant is calculated by summing the number density of all rovibrational energy levels: $n_{A}=\sum_{i}n_{i}$. The StS master equations were solved using the PLATO (Plasmas in thermodynamic non-equilibrium) library \cite{hindmarsh2005sundials,PLATO2023,Munafo_JTHT_plato_2025}. 

\subsection{Macroscopic quantities}
This section outlines the methodology used to determine the macroscopic quantities like recombination rate constants and internal energy by taking moments of the population distribution obtained by solving the system of master equations. The determination of these quantities is important as they may be compared against experimental data and used in the development of reduced-order models. 

\subsubsection{Effective recombination rate constant}
An effective recombination rate constant is computed based on the method proposed by Bourdon et. al \cite{Bourdon}. For a recombining environment, the dissociation source term on the right-hand side of Eq. (5) may be considered much smaller than the recombination rate term. Therefore, under a quasi-steady state condition, for a homogeneous chemical system, the effective recombination rate constant $\Bar{k}^{R}(T)$ may be defined as:
\begin{equation}
    \Bar{k}^{R}(T)=\frac{1}{2n_{B}^{3}}\left|\frac{dn_{B}}{dt}\right|.
\end{equation}

A global recombination rate constant is obtained from the sum of recombination source terms over all rovibrational energy levels at a particular time instant. This gives an expression for the global recombination rate constant $k_{g}^{R}(T)$ as a sum of all the microscopic recombination rate constants, obtained from dissociation rate constants using micro-reversibility, as follows:
\begin{align}
    k_{g}^{R}(T)=\sum k_{i}^{r}.
\end{align}
It is worth noting that the global recombination rate constant is independent of the population distribution at a given translational temperature. 

A QSS dissociation rate constant obtained from StS calculations is generally used in the development of reduced-order models \cite{Kim_Jo}. While this may hold well for dissociating environments, an in-depth analysis is required to determine its accuracy when applied to a recombining environment. Therefore, a third recombination rate constant $k^{R*}$ is calculated by imposing micro-reversibility on the QSS dissociation rate constant at a translational temperature. The macroscopic QSS dissociation rate constant is determined from the computed population of rovibrational energy levels for a dissociating case at a given heat-bath case and is expressed as \cite{Panesi2013_N3}:
\begin{equation}
    \Bar{k}^{D}(T)=\sum_{i}\frac{n_{i}k_{i}^{D}(T)}{n_{N_{2}}}.
\end{equation}
The macroscopic equilibrium constant is calculated using the formula:
\begin{equation}
    K_{eq}(T)=\frac{\left[g_{B}Q_{B}^{t}(T)\right]^{2}}{Q^{I}(T)Q_{A}^{t}(T)}exp\left(\frac{-E_{A}}{k_{B}T}\right).
\end{equation}
Micro-reversibility is then imposed to obtain the macroscopic recombination rate constant as follows:
\begin{equation}
    k^{R*}=\frac{\Bar{k}^{D}(T)}{K_{eq}(T)}.
\end{equation}

\subsubsection{Internal energy}
The molecular internal energy is computed by taking the first-order moment of the population distribution. The rovibrational energy is split into a rotational and a vibrational component. The vibrational energy pertaining to a particular vibrational quantum number is defined as the energy of the rotationless (J=0) level. The rotational energy is then calculated as the difference between internal and vibrational energies of the rovibrational level. The rotational and vibrational temperatures are calculated by solving implicit equations (Newton-Raphson method) which equate the concerned energy extracted from the computed population to an equivalent energy. The rotational and vibrational energies extracted from the population distribution are defined as:
\begin{equation}
    e_{A}^{R}=\sum_{v,J} \frac{n_{i(v,J)}E_{R}(v,J)}{n_{A}m_{A}},
\end{equation}
\begin{equation}
    e_{A}^{V}=\sum_{v}\frac{E_{V}}{m_{A}}\sum_{J}\frac{n_{i(v,J)}}{n_{A}},
\end{equation}
where $E_{V}$ and $E_{R}$ refer to the vibrational and rotational energy contribution of microscopic levels of vibrational and rotational quantum numbers v and J, respectively. The equivalent rotational and vibrational average energies,$e_{A}^{R*}$ and $e_{A}^{V*}$, respectively are defined as:
\begin{equation}
\begin{split}
    m_{A}e_{A}^{R*}(T^{R},T^{V})=\frac{1}{Q^{RV}(T^{R},T^{V})}\sum_{v}exp\left(\frac{-E_{V}}{k_{B}T^{V}}\right)\\
    \times\sum_{J}g_{i(v,J)}E_{R}(v,J)exp\left(\frac{-E_{R}(v,J)}{k_{B}T^{R}}\right),
\end{split}
\end{equation}

\begin{equation}
\begin{split}
    m_{A}e_{A}^{V*}(T^{R},T^{V})=\frac{1}{Q^{RV}(T^{R},T^{V})}\sum_{v}E_{V}exp\left(\frac{-E_{V}}{k_{B}T^{V}}\right)\\
    \times\sum_{J}g_{i(v,J)}exp\left(\frac{-E_{R}(v,J)}{k_{B}T^{R}}\right),
\end{split}
\end{equation}
The rovibrational partition function $Q^{RV}(T^{R},T^{V})$ is expressed as:
\begin{equation}
\begin{split}
    Q^{RV}(T^{R},T^{V})=\sum_{v}exp\left(\frac{-E_{V}}{k_{B}T^{V}}\right)\sum_{J}g_{i(v,J)}\\
    \times exp\left(\frac{-E_{R}(v,J)}{k_{B}T^{R}}\right).
\end{split}
\end{equation}

Further details about the methodology can be found in the work by Panesi et al \cite{Panesi2013_N3}.

\subsection{Reduced-order modeling} 
A previous study by Venturi et al. \cite{Venturi2020_O3Diss} proposed a hybrid approach for the master equation, which coupled the StS excitation with model reduction of the dissociation processes to determine the effectiveness of reduced-order models in accurately capturing the StS dissociation dynamics. A similar approach was used in the present study to analyze the different reduced-order models as applied to recombining environment. A second case, which uses coarse-grained modeling (CGM) of both dissociation and excitation processes, is also considered to analyze the effect of model reduction on the accurate capture of dissociation and excitation dynamics. The coarse-grained model relies on the binning of rovibrational levels into distinct bins, and the maximum-entropy principle is applied to the levels contained within each bin. This model reduction technique is described in detail by Liu et al \cite{Liu2015} and is briefly summarized here:
\begin{enumerate}
    \item The rovibrational levels are grouped into different bins
    \item The group-specific dissociation ($K_{P}^{D}$) and exchange ($K_{PQ}$) rate coefficients for bins P and Q are described as:\\
    \\
    $K_{P}^{D}=\sum_{i \in I_{P}} k_{i}^{d}(T)F_{i}^{p}$,       $K_{PQ}=\sum_{j \in I_{Q}}\sum_{i \in I_{P}} k_{ij}F_{i}^{p}$ \\
    where $F_{i}^{p}=\frac{g_{i}exp\left(\frac{-E_{i}}{k_{B}T}\right)}{Q^{I}(T)}$, $I_{P}$ and $I_{Q}$ represent energy levels contained in bins P and Q, respectively.
\end{enumerate}

In the first approach, the microscopic dissociation rate constants are reconstructed from the grouped dissociation rate constants. The reconstruction details for a homogeneous three-body system are detailed by Venturi et al \cite{Venturi2020_O3Diss}. In the work by Venturi et al, it was derived that the reconstructed rovibrational specific dissociation rate constants equal the group-specific dissociation rate constant. The derived dissociation rate constants and StS energy exchange kinetics are used to solve the StS master equations. 

For the second CGM case, the population of the P-th group is obtained by solving the coarse-grained master equations instead of the StS master equations:
\begin{equation}
    \frac{dn_{P}}{dt}=-K_{P}^{D}n_{P}n_{B}+K_{P}^{R}n_{B}^{3}-\sum_{Q \neq P}\left(K_{PQ}n_{P}n_{B}+K_{QP}n_{Q}n_{B}\right),
\end{equation}
\begin{equation}
    \frac{dn_{B}}{dt}=2\sum_{I_{P}}\left(K_{P}^{D}n_{P}n_{B}-K_{P}^{R}n_{B}^{3}\right).
\end{equation}
Here, $K_{P}^{R}$ refers to the group-specific recombination rate constant which is calculated by applying micro-reversibility to the group-specific dissociation rate constant. 

\section{RESULTS}
This section presents a detailed analysis of the StS kinetics involved in recombination and energy transfer processes when a chemical system is suddenly cooled to a low temperature in an isothermal and isochoric heat bath. The chemical systems, $N_{2}+N$ and $O_{2}+O$, primarily composed of atoms and a tiny fraction of molecules are initialized at a high temperature of 10,000 K and an initial pressure of $10^{5}$ Pa. This corresponds to an initial number density of $7.24 \times 10^{23} m^{-3}$. The initial mole fraction is taken to be the local thermodynamic equilibrium composition at initial conditions: $9.954 \times 10^{-1}$ mole fraction of nitrogen atoms and $4.6 \times 10^{-3}$ mole fraction of nitrogen molecules for the $N_{2}+N$ system and  $9.9997 \times 10^{-1}$ mole fraction of oxygen atoms and $3 \times 10^{-5}$ mole fraction of oxygen molecules for the $O_{2}+O$ chemical system. The system is then suddenly plunged into a low-temperature heat bath, thereby inducing non-equilibrium recombination and energy transfer processes. Two heat bath temperatures are chosen for the two chemical systems: the low-temperature case in which dissociation is almost negligible, and a high-temperature case in which molecules begin to dissociate.

%\subsection{$N_{2}+N$ chemical system}
\subsection{Study of dissociation/recombination processes}
\begin{figure}[h]
\includegraphics[width=0.5\textwidth]{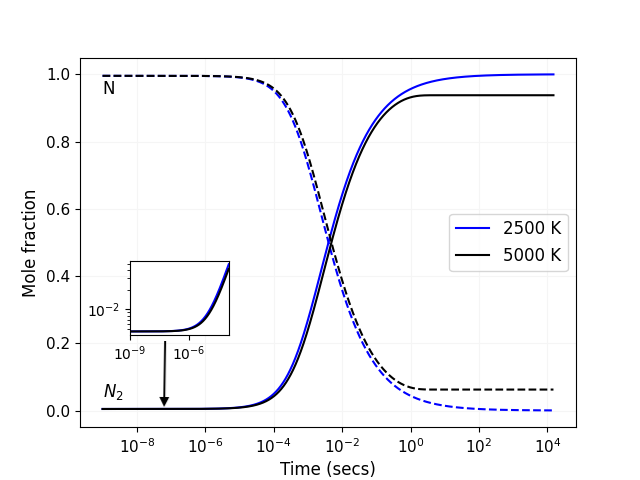}% Here is how to import EPS art
\caption{\textbf{Mole fraction evolution of N and $N_{2}$ at 2,500 K and 5,000 K.}}
\label{fig:N mole fraction}
\end{figure}

\begin{figure}
\includegraphics[width=0.5\textwidth]{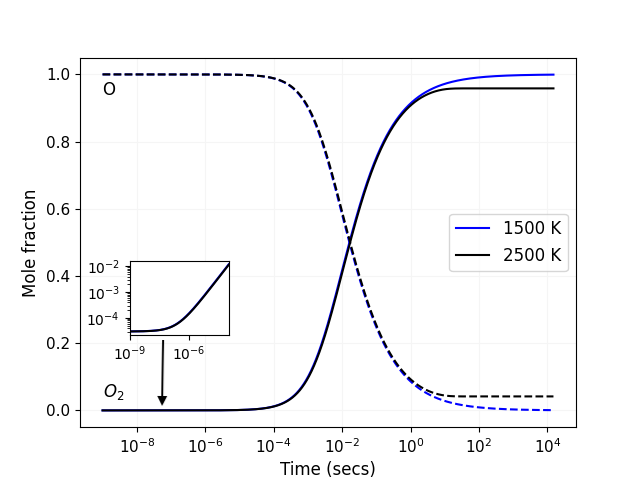}% Here is how to import EPS art
\caption{\textbf{Mole fraction evolution of O and $O_{2}$ at 1500 K and 2,500 K.}}
\label{fig:O mole fraction}
\end{figure}

\begin{figure*}
\centering
    \subfloat[]{\label{fig:s1}\includegraphics[width=0.5\textwidth]{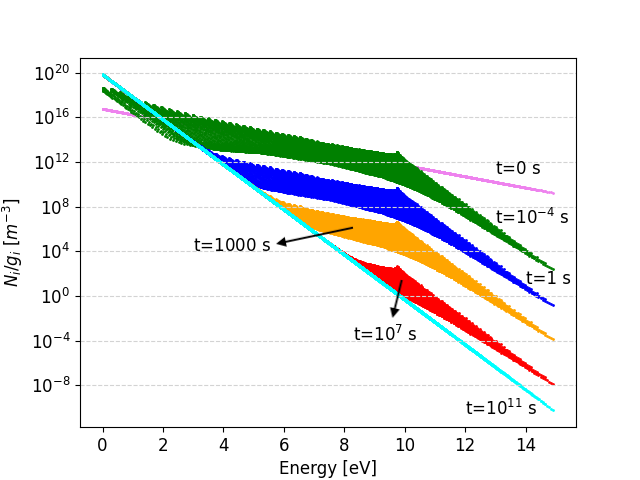}}
    \subfloat[]{\label{fig:s2}\includegraphics[width=0.5\textwidth]{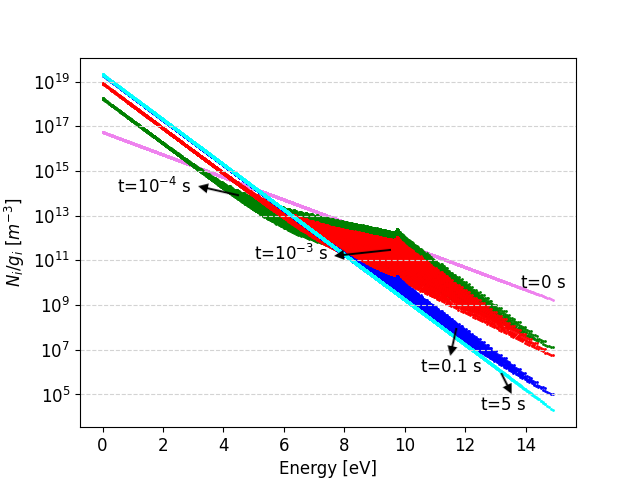}}
    \caption{\textbf{a) Population distribution evolution of $N_{2}$ at 2,500 K where violet=$0 s$, green=$10^{-4}s$, blue=$1s$, orange=$1000s$, red=$10^{7}s$, cyan=$10^{11}s$. b) Population distribution evolution of $N_{2}$ at 5,000 K where violet=$0 s$, green=$10^{-4}s$, red=$10^{-3}s$, blue=$0.1s$, cyan=$5s$}}
\end{figure*}

\begin{figure}
\includegraphics[width=0.5\textwidth]{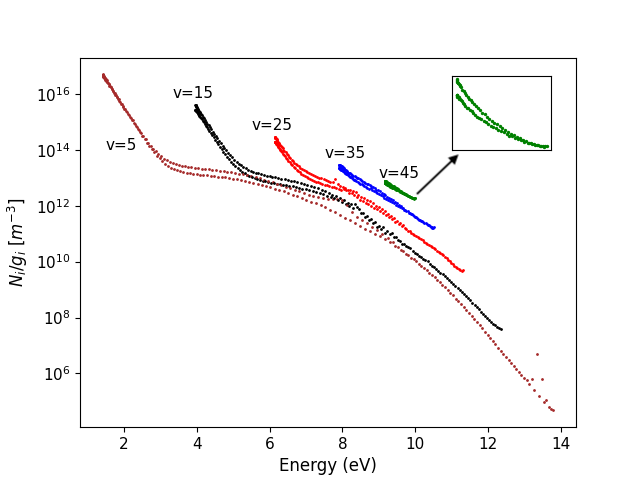}% Here is how to import EPS art
\caption{\textbf{Population dependence of $N_{2}$ molecules on vibrational quantum numbers at $t=10^{-6} s$ at 2,500 K heat bath temperature}}
\label{fig:N2 vqn}
\end{figure}

\begin{figure*}
\centering
    \subfloat[]{\label{fig:pop_O2_1500}\includegraphics[width=0.5\textwidth]{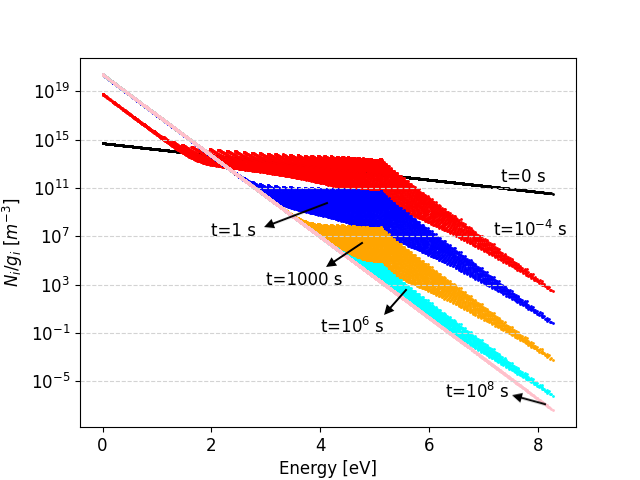}}
    \subfloat[]{\label{fig:pop_O2_2500}\includegraphics[width=0.5\textwidth]{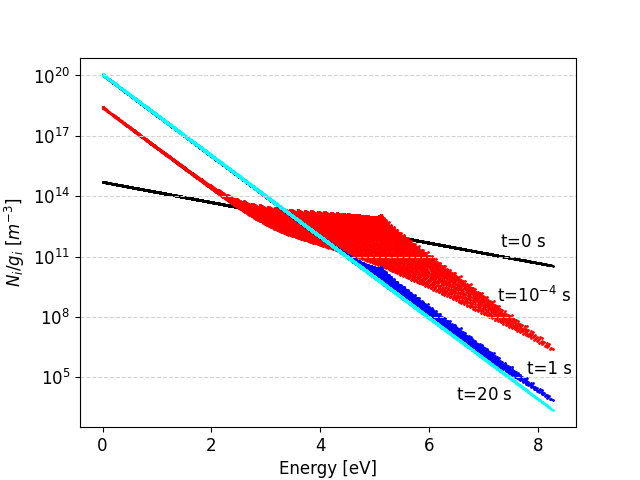}}
    \caption{\textbf{a) Population distribution evolution of $O_{2}$ at 1500 K where black=$0s$, red=$10^{-4}s$, blue=$1s$, orange=$1000s$, cyan=$10^{6}s$, pink=$10^{8}s$ b) Population distribution evolution of $O_{2}$ at 2,500 K where black=$0s$, red=$10^{-4} s$, blue= $1 s$, cyan=$20s$}}
\end{figure*}

The mole fraction evolution of nitrogen atoms and molecules for the two heat bath temperature cases of 2,500 K and 5,000 K is shown in Fig\ref{fig:N mole fraction}. The low-temperature heat bath experiences almost negligible dissociation, whereas dissociation effects become significant above 2,500 K. The bulk of recombination for both cases starts around the same time, shortly after $10^{-5}$ seconds; however, the atoms start recombining as early as $10^{-6}$ s, as shown in the inset figure above. There is a slight lag in the recombination observed at 5,000 K compared to 2,500 K, as the recombination rate is higher at lower temperatures, with dissociation being almost negligible. However, the chemical system at a higher heat bath temperature reaches chemical equilibrium within 5 seconds, whereas that at a lower heat bath temperature does not equilibrate until much later than 5 seconds. As can be seen from the figure, the bulk of recombination at 2,500 K is completed by $10^{2}$ secs, after which the change in mole fraction becomes gradual, suggesting a very slow recombination process thereafter. This is due to a reduction in the density of nitrogen atoms, thereby resulting in a reduced recombination rate. The difference in the evolution dynamics of the two temperatures may be attributed to the fact that dissociation is almost negligible at 2,500 K; therefore, in the absence of an opposing reaction, the chemical system evolves until all the atoms are used up. On the contrary, at 5,000 K, dissociation reactions come into play, and an early equilibrium stage is reached when the recombination and dissociation reactions balance each other. The equilibrium mole fraction of $N_{2}$ at 2,500 K ($\sim$1) is higher than that at 5,000 K ($\sim$0.94).

Similar to the recombination mechanism study of the $ N_2 + N$ chemical system, the mole fraction evolution of oxygen atoms and molecules for the two heat bath temperature cases of 1,500 K and 2,500 K is shown in Figure \ref{fig:O mole fraction}. The bulk of recombination for both cases begins around $10^{-4}$ seconds; however, the molecules start recombining as early as $10^{-7}$ seconds, as shown in the inset figure. The rate at which molecules recombine is very close for both the heat bath cases, in contrast to a slight lag observed in the nitrogen case. As observed in the low heat bath temperature case of $N_{2}+N$ system, the bulk of recombination at 1,500 K is completed by $10^{3}$ secs after which the rate of recombination becomes very slow. The equilibrium composition of oxygen molecules is $\sim 1$ at 1,500 K, showing complete recombination, while that at 2,500 K is a lower value of $\sim 0.96$. 

\begin{figure}[b]
\includegraphics[width=0.5\textwidth]{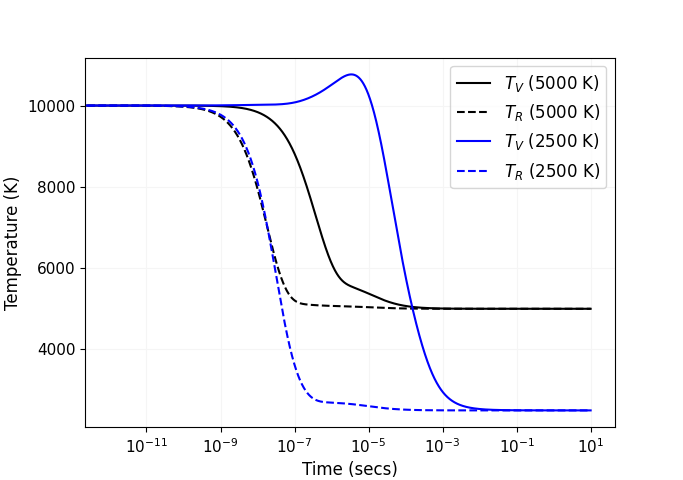}% Here is how to import EPS art
\caption{\textbf{Rotational and vibrational temperature evolution of $N_{2}$ at 2,500 K and 5,000 K.}}
\label{fig:N2 Temp}
\end{figure}

Figures \ref{fig:s1} and \ref{fig:s2} show the population distribution evolution of nitrogen molecules at heat bath temperatures of 2,500 K and 5,000 K, respectively. During the early stages of the simulation, the internal levels follow a Maxwell-Boltzmann distribution at the initial temperature of 10,000 K. The levels stay in the equilibrium distribution until $10^{-9}$ seconds for both cases, after which the distribution starts to deviate significantly. At this stage, the population evolution is characterized by distinct strands pertaining to a common vibrational quantum number. Each strand is in a state of partial equilibrium at a particular rotational temperature for low rotational quantum numbers. Figure \ref{fig:N2 vqn} illustrates the dependence of population distribution on vibrational quantum numbers at a time instant of $10^{-6} $ s for the low heat bath temperature case. The further splitting of vibrational strands into levels corresponding to odd and even rotational quantum numbers is observed for all vibrational levels with v > 5. The splitting is not observed in levels with high rotational quantum numbers, as shown in Figure \ref{fig:N2 vqn}. The vibration-specific partial equilibrium behavior is lost in the low vibrational energy strands towards the middle of the distribution. A similar trend was observed at a 5,000 K heat bath temperature.  A comparison of Figs \ref{fig:s1} and \ref{fig:s2} shows that the vibrational strands are more prominent at 2,500 K as compared to 5,000 K at $10^{-4}$ secs when the bulk of recombination commences. This may be explained by examining the internal temperature evolution as shown in Figure \ref{fig:N2 Temp}. By the time the bulk of recombination starts (\emph{i.e.}, $10^{-5}$ secs), the internal temperature has nearly thermalized at 5,000 K, resulting in a Maxwell-Boltzmann distribution of the low-lying energy states, while it is still far from the equilibrium value for the 2,500 K case. Therefore, the low-lying energy levels at 2,500 K take longer to reach equilibrium as compared to the 5,000 K heat bath temperature case, as shown by the green distributions in Figures \ref{fig:s1} and \ref{fig:s2}.  By studying the population evolution dynamics, it may be concluded that the dissociation/recombination and relaxation processes at 5,000 K are distinct from each other as most of the energy exchange occurs before the commencement of the bulk of recombination while the attainment of equilibrium distribution in the low-lying levels is delayed at 2,500 K due to the two processes occurring simultaneously.
The attainment of equilibrium in the high-lying vibrational states in both cases is delayed due to preferential recombination.

\begin{figure}
\includegraphics[width=0.5\textwidth]{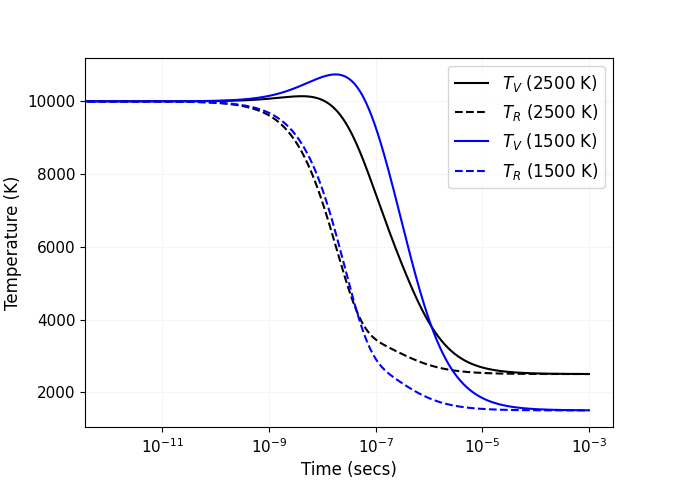}% Here is how to import EPS art
\caption{\textbf{Rotational and vibrational temperature evolution of $O_{2}$ at 1500 K and 2,500 K.}}
\label{fig:O2 Temp}
\end{figure}

The population distribution evolution of $O_{2}+O$ at 1,500 K and 2,500 K is shown in Figures \ref{fig:pop_O2_1500} and \ref{fig:pop_O2_2500}. The evolution is similar to that observed in $N_2 + N$, with an overpopulation of high-lying vibrational states, indicating a preferential recombination occurring in those states. It is also observed that at 1,500 K, the low-lying energy levels experience faster equilibration as compared to the low temperature heat bath case of 2,500 K  in $N_{2}+N$. It must be noted that the bulk of recombination commences by $10^{-4}$ s for both oxygen and nitrogen systems. Considering the population distribution at t=$10^{-4} s$, i.e, the instant at which the bulk of recombination commences, the results show equilibration of the low-lying energy levels up to approximately 1.5 eV.
In contrast, the population distribution of nitrogen molecules at t=$10^{-4} s$ as shown in Figure \ref{fig:s1} shows separation of low-lying energy levels in distinct vibrational strands. This may be attributable to the fact that, unlike the nitrogen system, the rotational and vibrational temperatures in the oxygen case are very close to reaching their equilibrium values by the onset of recombination, even at low temperatures, as shown in Figure \ref{fig:O2 Temp}. Therefore, the energy exchange processes resulting in the early relaxation of internal temperature lead to a Maxwell-Boltzmann distribution of the low-lying energy levels at the heat bath temperature.

Figure \ref{fig:N2 Temp} shows the time evolution of rotational and vibrational temperatures of nitrogen molecules at 2,500 K and 5,000 K heat bath temperatures starting from an initial internal temperature of 10,000 K. The rotational temperature relaxes at a much faster rate compared to the vibrational temperature, which is consistent with the conventional assumption that the rotational and translational timescales are comparable, with the vibrational relaxation timescale being higher at low temperatures. This leads us to the conclusion that using a common internal temperature when modeling the flows is not a valid assumption at low temperatures. It is also observed that the vibrational temperatures at 2,500 K begin to rise around $10^{-7}$ secs. This coincides with the time when recombination begins to occur, as shown in the inset plot of Figure \ref{fig:N mole fraction}. A plausible explanation for the temperature rise may be an increase in the population of high-lying vibrational states due to preferential recombination with the added effect of slow vibrational relaxation. A similar observation was made in the 5,000 K case, where the rate of decrease in vibrational temperature reduces around $10^{-6}$ seconds when recombination starts. Therefore, a coupling between the energy relaxation and dissociation/recombination processes delays the thermalization process. The effect of dissociation/recombination processes is more pronounced on the evolution of vibrational temperature compared to the rotational temperature, indicating a stronger chemical-vibrational coupling that results in the addition of more energy to the vibrational mode during the production of nitrogen molecules. Figure \ref{fig:O2 Temp} shows the evolution of vibrational and rotational temperature for the $O_{2}+O$ non-equilibrium recombination process. As the figure shows, the relaxation time scales for the two heat bath temperatures are very close, as compared to a considerable disparity observed in the nitrogen system. Similar to the nitrogen system, the preferential recombination of high-lying vibrational states results in an increase in vibrational temperature at 1,500 K, around $10^{-8}$ s, which is the time instant when recombination commences and the mole fraction of molecules starts to increase. 

\subsection{Study of energy transfer processes}
\begin{figure*}
\centering
    \subfloat[]{\label{fig:s11}\includegraphics[width=0.5\textwidth]{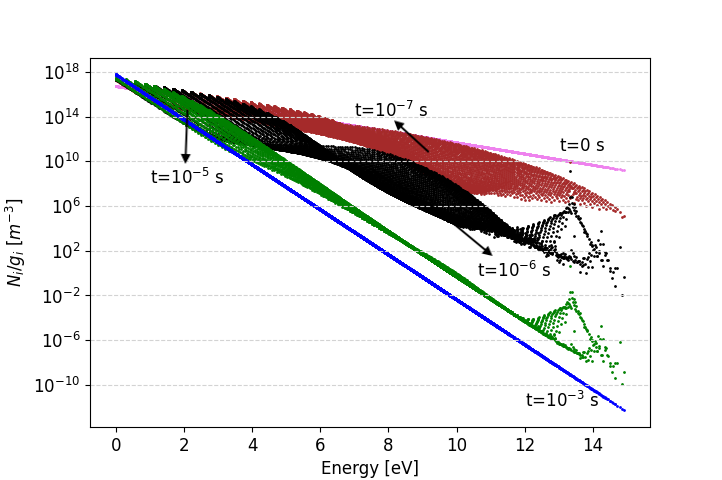}}
    \subfloat[]{\label{fig:s21}\includegraphics[width=0.5\textwidth]{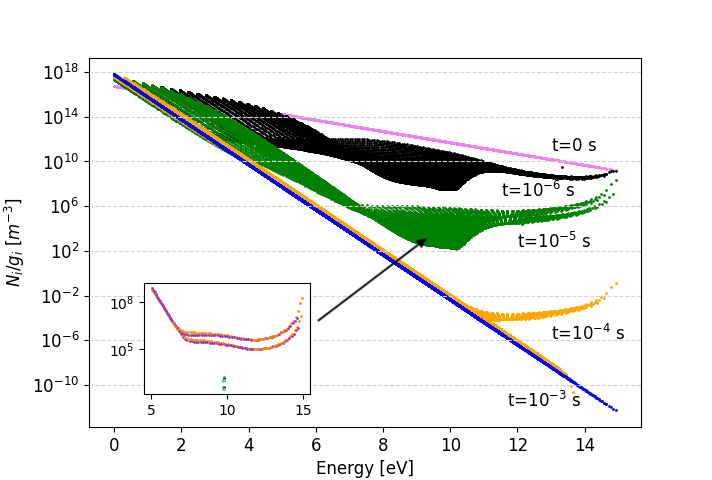}}
    \caption{\textbf{a) Population distribution evolution of $N_{2}$ at 2,500 K with exchange and inelastic processes included where violet=$0 s$, brown=$10^{-7}s$, black=$10^{-6}s$, green=$10^{-5}s$, blue=$10^{-3}s$ b) Population distribution evolution of $N_{2}$ at 2,500 K considering only inelastic processes where violet=$0 s$,black=$10^{-6}s$, green=$10^{-5}s$,orange=$10^{-4}s$, blue=$10^{-3}s$}}
\end{figure*}

This section aims to study the dynamics of exchange and inelastic processes in non-equilibrium cooling of $N_{2}+N$ and $O_{2}+O$ chemical systems by excluding the consideration of dissociation/recombination reactions. Some earlier studies \cite{Tau_Expanding1,Tau_Expanding2,Tau_Expanding3,Tau_Expanding4,Tau_Expanding5,Tau_Expanding6} focused on determining whether the vibrational relaxation time in expanding flows is similar to that determined from the dissociating shock wave experiments. The availability of StS data has now enabled the numerical quantification of the energy relaxation phenomenon, thereby better informing the reduced-order models. The initial conditions are the same as those for the study of dissociation/recombination processes. Only the cases with low heat bath temperatures are discussed in detail for both chemical systems.

Figure \ref{fig:s11} shows the population distribution evolution of nitrogen molecules when both exchange and inelastic processes are considered, while Figure \ref{fig:s21} considers only inelastic processes. Initially, the rovibrational levels exhibit a Maxwell-Boltzmann distribution at an initial internal temperature of 10,000 K. With time, the collisional processes result in the transfer of energy to the rotational and vibrational modes. The low-lying energy levels achieve a partial equilibrium stage, as observed in the dissociation/recombination section.
In contrast, the dynamics of energy levels above 6 eV are significantly different from the recombination case due to the absence of dissociation/recombination kinetics, which dominate that regime of energy. Figure \ref{fig:s21} shows that the inelastic processes result in an overpopulation of the high-lying energy levels, with the separation of odd and even rotational quantum numbers into two distinct strands, as observed in the inset figure, where different colors represent vibrational quantum numbers. However, this behavior was not observed in the study of inelastic processes at 5,000 K, where the results showed no further splitting of the vibrational levels into distinct strands of odd and even rotational quantum numbers in the high-lying energy levels. It is also important to note here that quantum mechanics forbids inelastic transitions between odd and even rotational quantum numbers. This effect is not very prominent when exchange processes are included, as shown in Figure \ref{fig:s11}. The exchange processes can result in a broad spectrum of energy exchange reactions by replacing one of the bound atoms of the nitrogen molecule with that of the colliding partner, thereby reducing the effect of selection rules observed in inelastic processes alone. The exchange reactions promote the thermalization process of the high-lying energy levels and the quasi-bound levels by making jumps of multiple vibrational levels more probable, as shown at t=$10^{-5} s$ in Figures \ref{fig:s11} and \ref{fig:s21}. The presence of an energy exchange barrier for the low-lying energy levels delays their thermalization to a common internal temperature \cite{Panesi2013_N3}. The population for exchange+inelastic and inelastic cases reaches the Maxwell-Boltzmann distribution at the heat bath temperature of 2,500 K around the same instant of time at t=$10^{-3}$s. These results indicate that the low-lying energy levels have the highest impact on relaxation time due to their slow thermalization process. 

\begin{figure*}
\centering
    \subfloat[]{\label{pop_O2_EI}\includegraphics[width=0.5\textwidth]{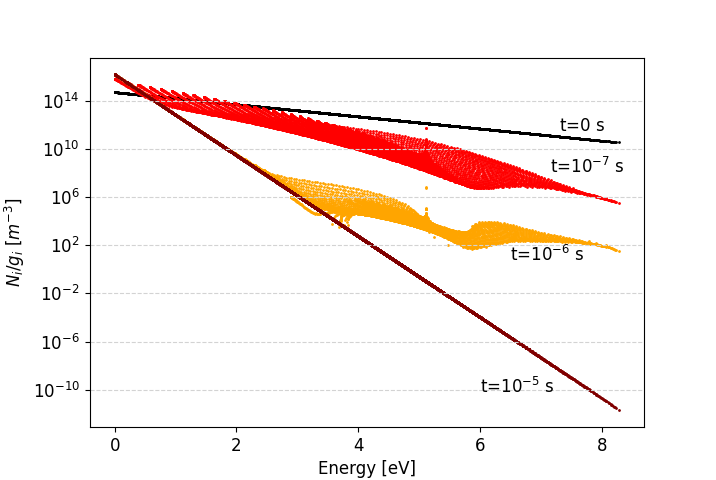}}
    \subfloat[]{\label{pop_O2_Inel}\includegraphics[width=0.44\textwidth]{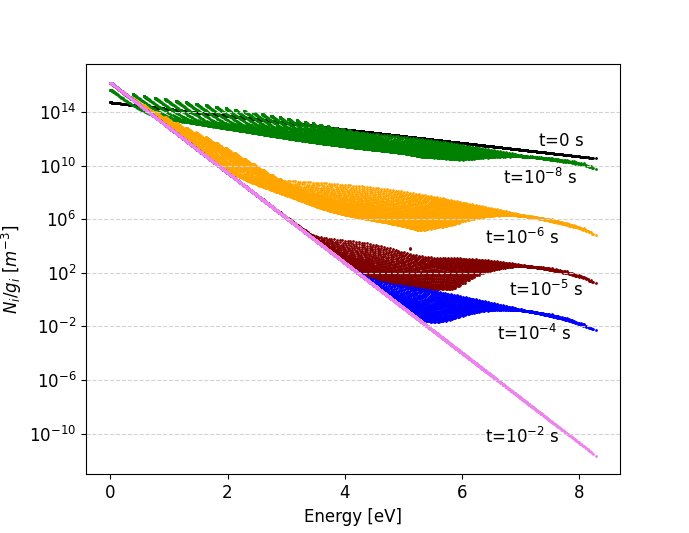}}
    \caption{\textbf{a) Population distribution evolution of $O_{2}$ at 1500 K with exchange and inelastic processes included where black=$0s$, green=$10^{-8}s$, red=$10^{-7}s$, orange=$10^{-6}s$, maroon=$10^{-5}s$  b) Population distribution evolution of $O_{2}$ at 1500 K considering only inelastic processes where black=$0 s$, red=$10^{-8}s$, blue=$10^{-6}s$, maroon=$10^{-5}s$, orange=$10^{-4}s$,violet=$10^{-2} s$}}
    
\end{figure*}

\begin{figure}
\includegraphics[width=0.5\textwidth]{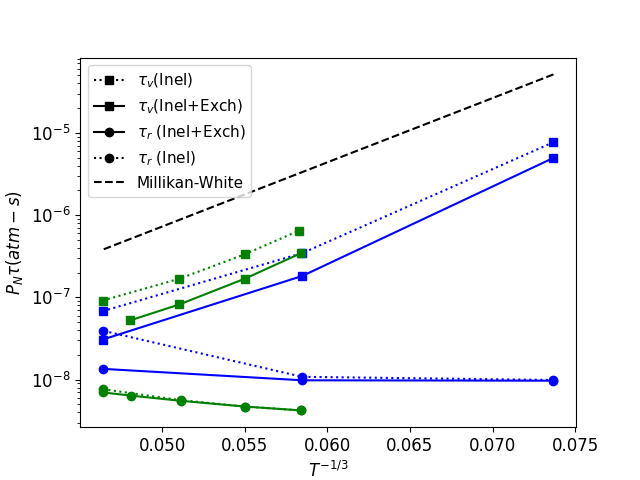}% Here is how to import EPS art
\caption{\textbf{Rotational and vibrational relaxation times of $N_{2}$ as a function of translational temperature. Green color shows heating case results obtained from Panesi et al. Blue color shows cooling results of the current study}}
\label{fig:N2 Relax}
\end{figure}

A study of the energy exchange processes concerning $O_{2}+O$ chemical system is presented in this paragraph. The population distribution evolution at 1,500 K for inelastic plus exchange and inelastic processes alone is shown in Figure \ref{pop_O2_EI} and Figure \ref{pop_O2_Inel}, respectively. At the early stages of simulation, the low-lying energy levels separate into distinct vibrational strands at a common rotational temperature close to the heat bath temperature as observed in nitrogen case. As opposed to what was observed in the analysis of energy transfer in $N_{2}+N$, the low-lying levels of oxygen equilibrate much earlier than the high-lying energy states as the system approaches equilibrium. This indicates the absence of a high energy exchange barrier in the low-lying states as present in the nitrogen case. The high-lying energy levels in the case considering inelastic processes alone exhibit overpopulation with the internal temperature being closer to the initial temperature. In the absence of exchange reactions, the system takes much longer to reach thermodynamic equilibrium as compared to when exchange is taken into consideration. The disparity in relaxation for the two cases is higher than that observed in nitrogen, where the chemical system reached equilibrium at almost the same time, regardless of whether exchange is included or not.

\begin{figure}[h]
\includegraphics[width=0.5\textwidth]{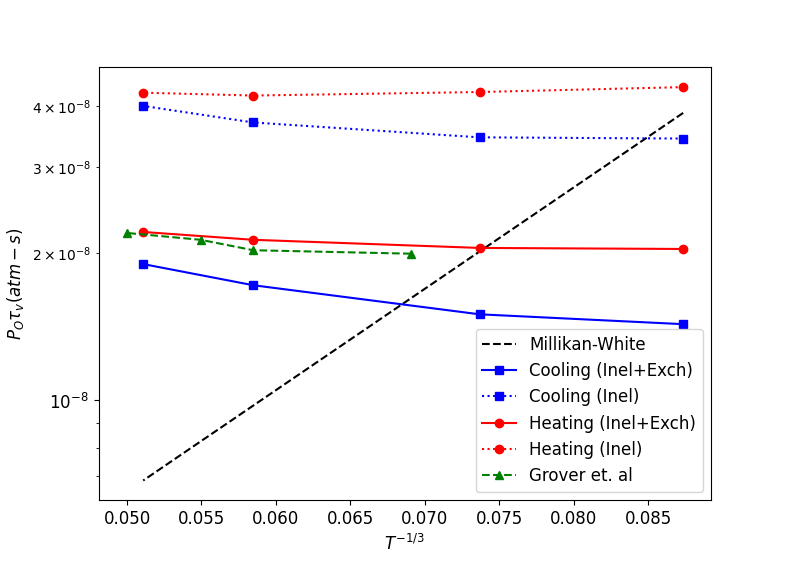}% Here is how to import EPS art
\caption{\textbf{Vibrational relaxation times of $O_{2}$ as a function of translational temperature.}}
\label{fig:O2 Relax hc}
\end{figure}

\begin{figure}
\includegraphics[width=0.5\textwidth]{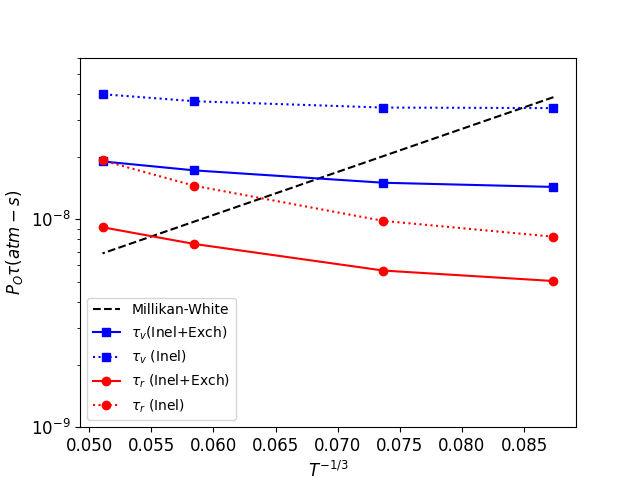}% Here is how to import EPS art
\caption{\textbf{Rotational and vibrational relaxation times of $O_{2}$ as a function of translational temperature.}}
\label{fig:O2 Relax rv}
\end{figure}

Further insights into the relaxation process are gained by examining the rotational and vibrational relaxation times and energy evolution of nitrogen. Figure \ref{fig:N2 Relax} shows the rotational and vibrational relaxation times for heating and cooling conditions. The cooling conditions are created by suddenly cooling the gas mixture from an initial high temperature to a low-temperature heat bath, and considering only energy transfer processes. The heating case, on the other hand, starts from a low initial temperature, and the system is suddenly heated to a very high temperature without considering dissociation reactions. A comparison of relaxation times obtained using inelastic and exchange processes, and inelastic processes alone, is carried out to understand the influence of exchange reactions. The results of the present study for the cooling case are shown in blue, while the heating results, obtained from Panesi et al. \cite{Panesi2013_N3}, are shown in green. A comparison of the vibrational relaxation times obtained for heating and cooling cases shows that the product $P_{N} \tau$ is lower for cooling cases, while the rotational relaxation time constant $P_{N} \tau$ is higher in cooling environments. A difference in the vibrational relaxation times observed in the heating and cooling cases may be explained by the effect of anharmonicity in expanding flows \cite{Park}. The high-lying vibrational levels in the heating case are less populated than the equilibrium value, while they are overpopulated in the cooling case. These high-lying vibrational levels, due to their anharmonic nature, introduce anharmonicity in the cooling flows, thereby influencing the relaxation time. The inclusion of exchange processes lowers the relaxation time for both vibrational and rotational modes, with the effect on rotational mode being almost negligible at low temperatures. The relaxation time scale difference between vibrational and rotational modes decreases as the temperature increases. A comparison of vibrational relaxation times obtained from master equations to those obtained from the Millikan-White correlations \cite{Millikan-White} shows a similar temperature dependence; however, there is a nearly order-of-magnitude difference between the two, with the difference being even higher for the cooling case.

\begin{table*}
\begin{centering}
\begin{tabular}{ | m{3cm} | m{3cm}| m{3cm} | m{3cm} |} \hline
Chemical system & Temperature (K) & $\phi(P_{B}\tau_{v})$ & $\phi(P_{B}\tau_{r})$ \\
\hline
$N_{2}+N$ & 2,500 & 5.25 & 1.02\\
    &5,000 & 1.32 & 1.01 \\
    \hline
$O_{2}+O$ & 1,500 & 6.52 & 1.06 \\
    &2,500 & 5.36 & 1.08\\
    \hline
\end{tabular}
\caption{Ratio between relaxation time constant for vibrational and rotational modes}
\label{tab:my_label1}
\end{centering}
\end{table*}

An analysis of relaxation time, including recombination processes, was carried out and compared with energy transfer processes (inelastic + exchange) alone to quantify the coupling between recombination and energy transfer phenomena. A parameter $\phi$ is defined as the ratio between $P_{B} \tau_{r,v}$ values for simulations with and without recombination processes included. The results shown in Table \ref{tab:my_label1} indicate that the ratio between the two values is higher for vibrational modes, whereas it is nearly unity for rotational modes. This signifies a coupling between energy transfer and recombination processes at low temperatures, with the effect being more prominent on the vibrational mode.

 The relaxation time values for the recombining case in oxygen are shown in Figure \ref{fig:O2 Relax rv}. At 1,500 K, the ratio of vibrational relaxation time within inelastic processes to inelastic plus exchange processes is 2.4, and that of rotational relaxation time for the two cases is 1.63. On the other hand, at the low heat bath temperature case of 2,500 K for nitrogen, the ratio of vibrational relaxation times is 1.5 and that of rotational relaxation times is 1. Similar to the nitrogen system, Figure \ref{fig:O2 Relax hc} shows that the relaxation time is different for heating and cooling cases. However, the difference in the two is not as substantial as that observed in nitrogen. It is also observed that the dependence of relaxation times on temperature ($T^{-1/3}$) does not follow the same trend as Millikan-White correlation for $O_{2}+O$ \cite{Boyd_O21} even in the cooling conditions. The results obtained for the heating case, considering both inelastic and exchange processes, agree well with those published by Grover et al. \cite{Grover_tau}, which used the same ab initio PESs.

\begin{figure}
\includegraphics[width=0.5\textwidth]{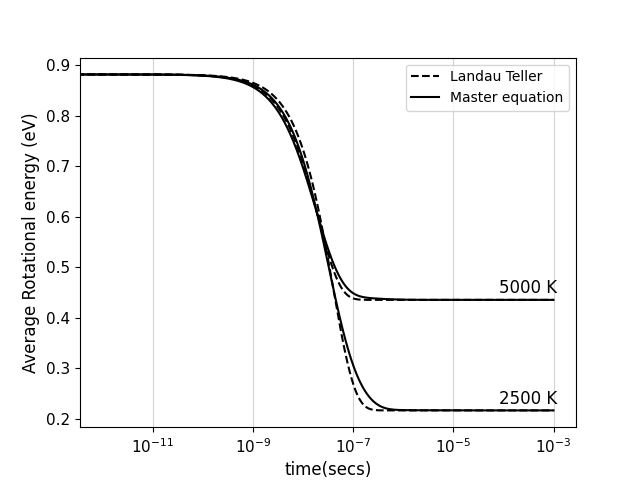}% Here is how to import EPS art
\caption{\textbf{Comparison of averaged rotational energy obtained from master equation and Landau-Teller relaxation}}
\label{fig:N2 Rot}
\end{figure}

\begin{figure}
\includegraphics[width=0.5\textwidth]{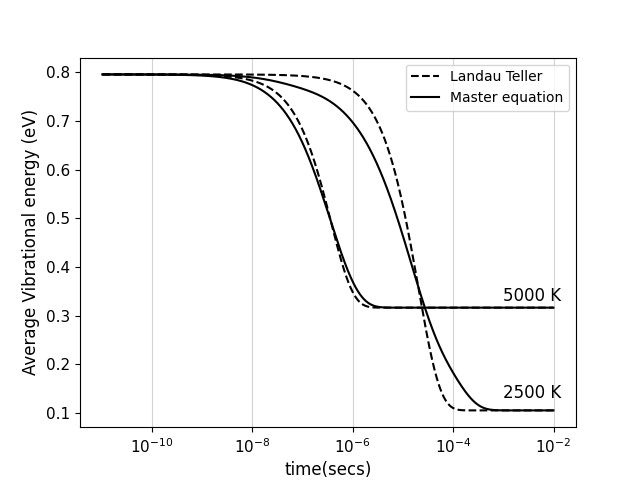}% Here is how to import EPS art
\caption{\textbf{Comparison of averaged vibrational energy obtained from master equation and Landau-Teller relaxation}}
\label{fig:N2 Vib}
\end{figure}

\begin{figure}
\includegraphics[width=0.5\textwidth]{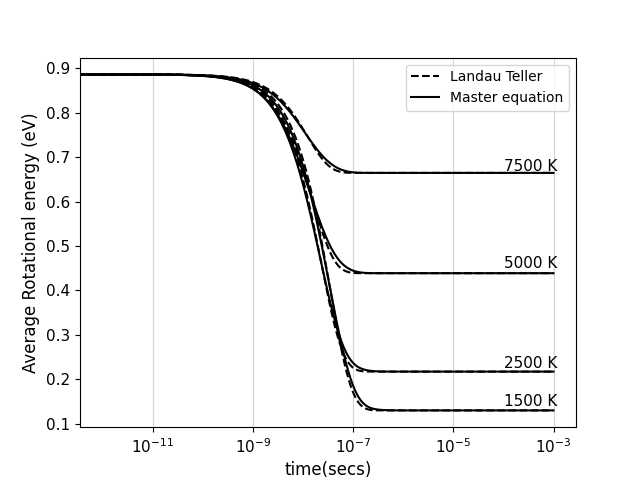}% Here is how to import EPS art
\caption{\textbf{Comparison of averaged rotational energy obtained from master equation and Landau-Teller relaxation}}
\label{fig:O2 Rot}
\end{figure}

\begin{figure}[h]
\includegraphics[width=0.5\textwidth]{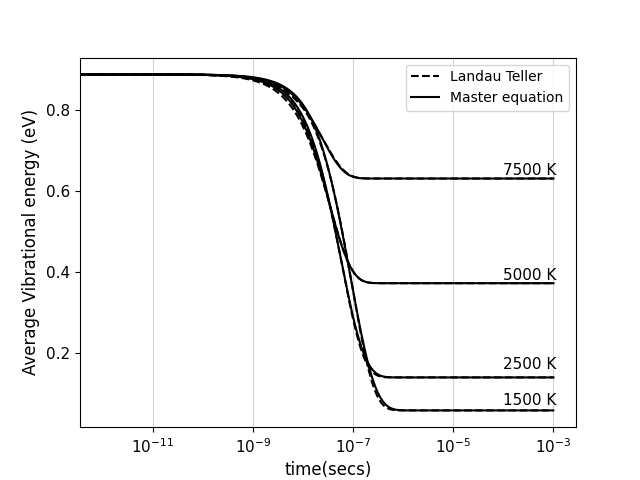}% Here is how to import EPS art
\caption{\textbf{Comparison of averaged vibrational energy obtained from master equation and Landau-Teller relaxation}}
\label{fig:O2 Vib}
\end{figure}

Figures \ref{fig:N2 Rot} and \ref{fig:N2 Vib} show a comparison of the evolution of average rotational and vibrational energies, respectively, obtained from the master equation and Landau-Teller relaxation for nitrogen. A look at the figures reveals that the rotational energy evolution is captured very well by the Landau-Teller relaxation model; however, this is not the case with vibrational energy relaxation, especially at low temperatures. Figure \ref{fig:N2 Vib} shows a considerable difference in the evolution plot at 2,500 K suggesting that Landau-Teller relaxation may not be a good approximation for vibrational relaxation with a significant degree of non-equilibrium at 2,500 K. A similar comparison of the evolution of rotational and vibrational energies for the $O_{2}+O$ system is shown in Figures \ref{fig:O2 Rot} and \ref{fig:O2 Vib} respectively. It can be inferred that for the oxygen case, the Landau-Teller model accurately captures the energy evolution at all temperatures, with the accuracy being higher for vibrational energy. 

\subsection{Macroscopic recombination rate constant}
\begin{figure}
\includegraphics[width=0.5\textwidth]{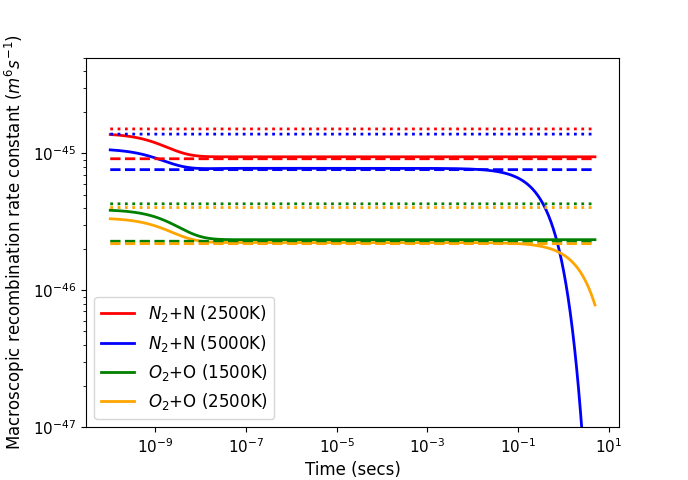}% Here is how to import EPS art
\caption{\textbf{Macroscopic recombination rate constant for $N_{2}+N$ and $O_{2}+O$ where dotted lines indicate $k_{g}^{R}$,solid lines indicate $\bar{k}^{R}$ and dashed lines indicate $k^{R*}$}}
\label{fig:rate constant}
\end{figure}

\begin{table*}
\begin{center}
\begin{tabular}{ | m{3cm} | m{3cm}| m{2cm} | m{2cm} | m{2cm} |} \hline
Chemical system & Temperature (K) & $k_{g}^{R} (m^{6} s^{-1})$ & $\bar{k}^{R} (m^{6}s^{-1})$ & $k^{R*} (m^{6}s^{-1})$ \\
\hline
$N_{2}+N$ & 2,500 & $1.507 \times 10^{-45}$ & $9.125 \times 10^{-46}$ & $9.43 \times 10^{-46}$\\
    &5,000 & $1.38 \times 10^{-45}$ & $7.6 \times 10^{-46}$ & $7.77 \times 10^{-46}$\\
    \hline
$O_{2}+O$ & 1500 & $4.277 \times 10^{-46}$ & $2.341 \times 10^{-46}$ & $2.28 \times 10^{-46}$\\
    &2,500 & $4.018 \times 10^{-46}$ & $2.232 \times 10^{-46}$ & $2.19 \times 10^{-46}$\\
    \hline
\end{tabular}
\caption{Macroscopic recombination rate constant values}
\label{tab:my_label}
\end{center}
\end{table*}

\begin{figure*}
\centering
    \subfloat[]{\label{fig:N3 diss binning}\includegraphics[width=0.5\textwidth]{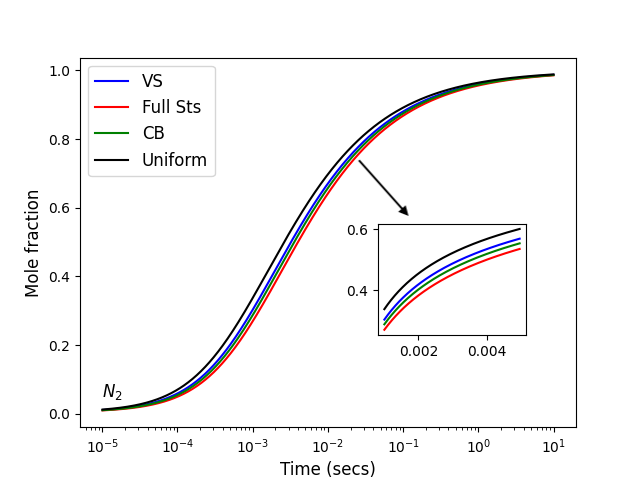}}
    \subfloat[]{\label{fig:O3 diss binning}\includegraphics[width=0.5\textwidth]{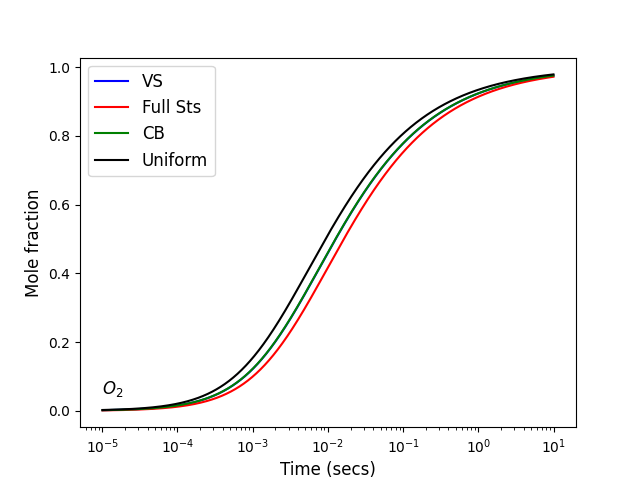}}
    \caption{\textbf{Mole fraction evolution for a) $N_{2}+N$ (2,500 K) and b) $O_{2}+O$ (1500 K) using 10 groups vibration-specific (VS), distance from centrifugal barrier based approach (CB) and Uniform energy binning}}
\end{figure*}

\begin{figure*}
\centering
    \subfloat[]{\label{fig:error_diss_N2}\includegraphics[width=0.5\textwidth]{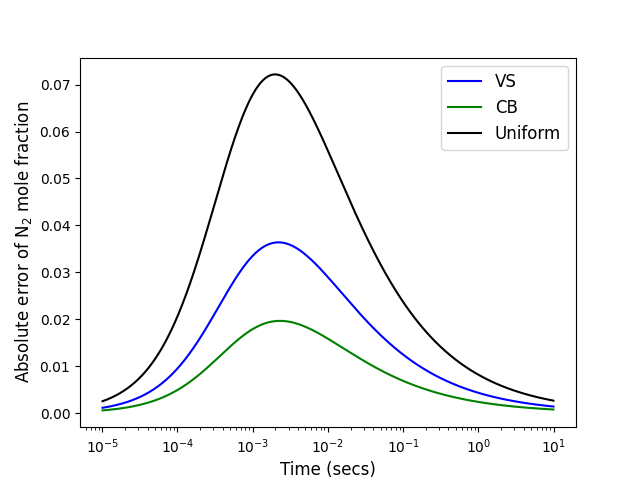}}
    \subfloat[]{\label{fig:error_diss_O2}\includegraphics[width=0.5\textwidth]{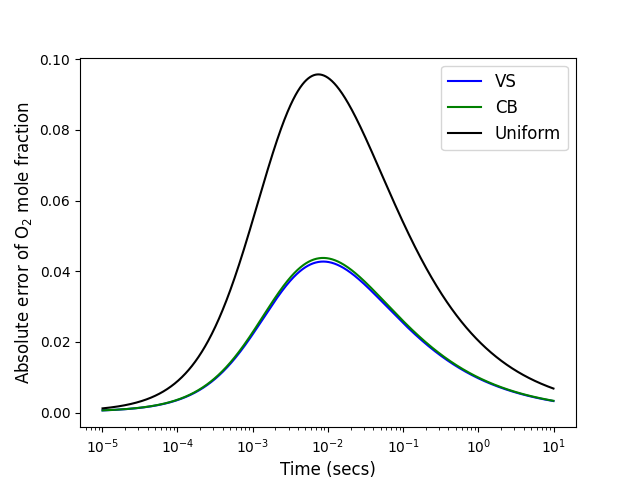}}
    \caption{\textbf{Absolute error of mole fraction for a) $N_{2}+N$ (2,500 K) and b) $O_{2}+O$ (1500 K) using 10 groups vibration-specific (VS), distance from centrifugal barrier based approach (CB) and Uniform energy binning}}
\end{figure*}

\begin{figure*}
\centering
    \subfloat[]{\label{fig:N3 de binning}\includegraphics[width=0.5\textwidth]{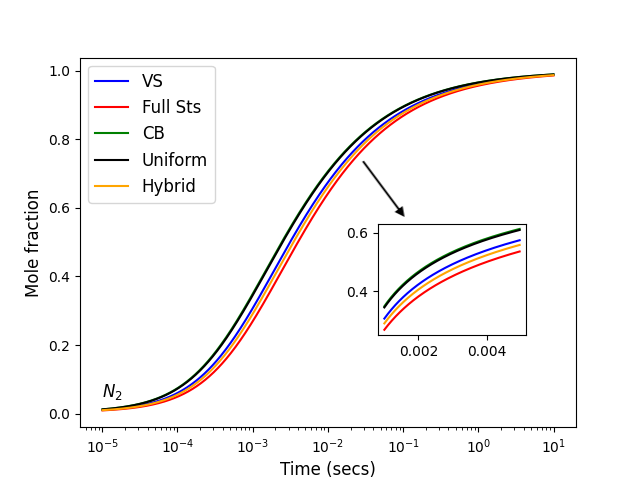}}
    \subfloat[]{\label{fig:O3 de binning}\includegraphics[width=0.5\textwidth]{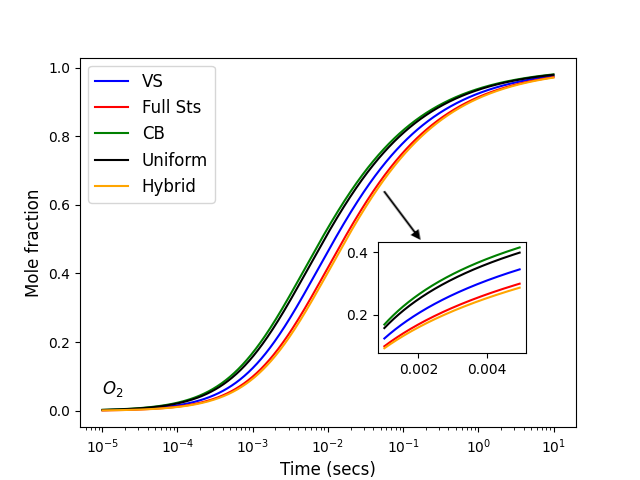}}
    \caption{\textbf{Mole fraction evolution for a) $N_{2}+N$ (2,500 K) and b) $O_{2}+O$ (1500 K) using 10 groups vibration-specific (VS), adaptive, distance from centrifugal barrier based approach (CB) and Uniform energy binning}}
\end{figure*}

\begin{figure*}
\centering
    \subfloat[]{\label{fig:error_N2}\includegraphics[width=0.5\textwidth]{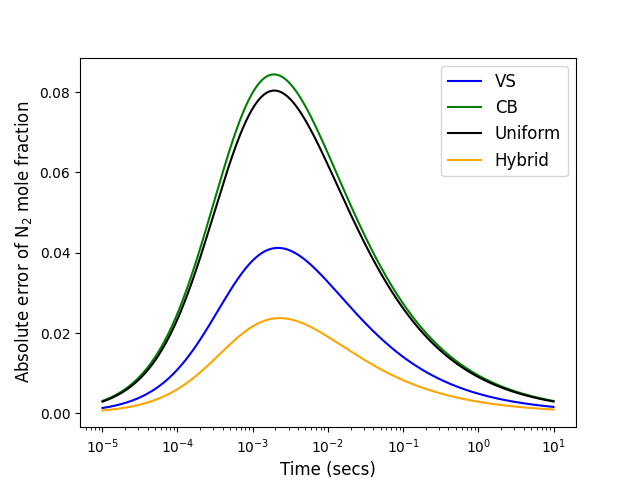}}
    \subfloat[]{\label{fig:error_O2}\includegraphics[width=0.5\textwidth]{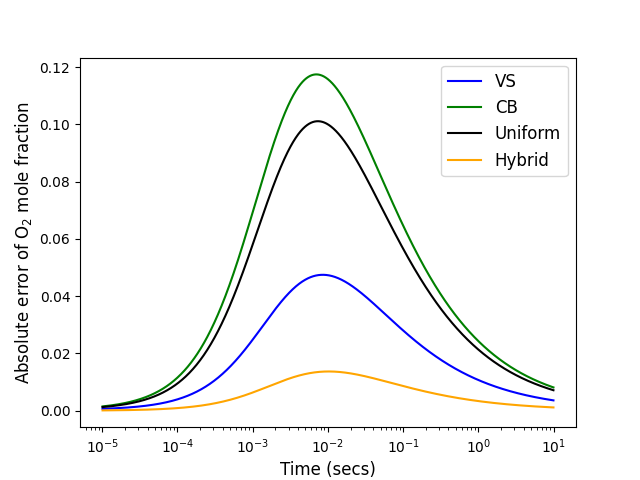}}
    \caption{\textbf{Absolute error of mole fraction for a) $N_{2}+N$ (2,500 K) and b) $O_{2}+O$ (1500 K) using 10 groups vibration-specific (VS), adaptive, distance from centrifugal barrier based approach (CB) and Uniform energy binning}}
\end{figure*}

A comparison of the macroscopic recombination rate constants $k_{g}^{R}$, $\bar{k}^{R}$ and $k^{R*}$ obtained using different approaches as described in Section II D is presented in Figure \ref{fig:rate constant} for the two chemical systems under consideration. The result shown in Figure \ref{fig:rate constant} suggests that $\bar{k}^{R}$ and $k^{R*}$ give the same value for the effective recombination rate constant. Therefore, the QSS dissociation rate constant obtained from Full StS analysis may be used to compute the QSS recombination rate constant for application in the development of reduced-order models. There is a considerable difference observed between the recombination rate constants obtained from the master equation and the global recombination rate constant $k_{g}^{R}$. A plausible explanation for this difference is that, although negligible, dissociation is considered while simulating the 0-D StS cases. The two recombination rate constants differ by a factor of 1.65 for $N_{2}+N$ 2,500 K case, 1.81 for $N_{2}+N$ 5,000 K case, 1.87 for $O_{2}+O$ 1,500 K case and 1.83 for $O_{2}+O$ 2,500 K case. The effective macroscopic recombination rate constant values are tabulated in Table \ref{tab:my_label}.

\subsection{Assessment of accuracy of binning strategies for recombining cases}
\subsubsection{CGM applied to dissociation kinetics only}
This section aims to assess the accuracy of multiple binning strategies by applying CGM to chemical kinetics and comparing the reduced-order model results with the Full StS. The results presented in the following sections are based on low heat bath temperatures of 2,500 K and 1,500 K for nitrogen and oxygen, respectively.

This section presents the results of coarse-grained modeling for 10 groups, applied only to the dissociation kinetics, while considering a Full StS resolution of the energy transfer processes. The grouping strategies considered here are:

\begin{enumerate}
    \item Uniform grouping: The rovibrational levels are grouped into 10 bins by dividing the internal energy into equally distributed segments.
    \item Vibration-specific (VS): The vibrational quantum numbers are equally divided into 10 intervals.
    \item Distance from centrifugal-barrier based (CB): The levels are clustered based on the energy deficit from the centrifugal barrier \cite{Venturi2020_O3Diss}. The grid is not uniform and is refined in the proximity of the centrifugal barrier up to an energy deficit of 1 eV for both chemical systems. 
\end{enumerate}

Figures \ref{fig:N3 diss binning} and \ref{fig:O3 diss binning} show the mole fraction evolution for the two chemical systems employing the above-mentioned binning strategies. The results show that the binning strategies work similarly for both systems, with uniform grouping showing the most significant discrepancy when compared to StS results. The VS and CB grouping strategies give overlapping results for oxygen, while CB performs better than VS for nitrogen. This result for the recombining case is also in agreement with the study conducted by Venturi et al. \cite{Venturi2020_O3Diss}, which concluded that CB works much better than VS in a dissociating environment. Although CB performs better or the same as VS for recombination in the two systems, the difference is not as significant as observed in the dissociation case. To gain a better quantitative understanding of the accuracy, Figures \ref{fig:error_diss_N2} and \ref{fig:error_diss_O2} display the magnitude of the absolute error observed in mole fraction evolution when employing different binning strategies for the recombination process. It must be noted here that the outcome of the CB strategy is highly dependent on the refinement of clustering in the proximity of the centrifugal barrier. Further analysis is needed to determine the optimal energy deficit from the centrifugal barrier to achieve the best results. 

\subsubsection{CGM applied to dissociation and energy transfer kinetics}
CGM is applied to both dissociation and energy transfer kinetics to determine an accurate binning strategy for the complete recombination phenomenon. The binning strategies considered are the same as those considered in the previous section, with the addition of a hybrid strategy. The hybrid binning strategy is a combination of CB grouping in the high-lying energy levels and network clustering \cite{Venturi2020_O3Diss} in the low-lying energy levels, where energy exchanges are prominent. The network clustering is performed using the Python network clustering algorithm, Infomap, on the inelastic kinetics. The results in Figures \ref{fig:N3 de binning} and \ref{fig:O3 de binning} show that while VS binning works better than uniform grouping for both chemical systems, CB grouping fails to accurately capture the dynamics. This leads us to conclude that CB is not suitable for modeling energy exchange processes, especially at low-lying energy levels where energy transfer is dominant. A network clustering method that groups the energy levels with intense energy exchanges among them, and hence highly likely to equilibrate with each other, is required to model the energy transfer processes. To this end, the hybrid strategy is proposed, which outperforms all the other binning strategies. Figures \ref{fig:error_N2} and \ref{fig:error_O2} show the magnitude of absolute error observed in mole fraction evolution when employing different binning strategies for recombination and energy transfer processes, with the hybrid strategy being the most accurate.

\section{Summary and Conclusions}
The evolution of population distribution and macroscopic quantities like mole fraction, internal temperature, recombination rate constant, and relaxation time constants were determined for the non-equilibrium recombining conditions for $ N_2 +N$ and $ O_2 +O$ chemical systems in a 0-D isothermal chemical reactor. The rotational temperature relaxes at a much faster rate than the vibrational temperature for nitroge,n while the relaxation times are comparable for oxygen. The numerical studies also show a coupling between the recombination and energy transfer processes, with the two occurring simultaneously during part of the simulation for the low-temperature heat bath case of nitrogen, while they are distinct for the other cases considered. This results in a more prominent vibrational strand-like structure of the low-lying energy levels when the recombination commences at 2,500 K for nitrogen. The recombination phenomenon is characterized by an overpopulation of the high-lying vibrational states, indicating preferential recombination occurring there. It was observed that the population distribution evolution is highly vibration-specific except in the middle of the distribution. The rotational and vibrational relaxation times were found to differ between heating and cooling cases at a given translational temperature, with the relaxation time constant being lower for the cooling case due to the anharmonic effect of the high-lying vibrational states. An attempt was also made at model reduction of the simulation conditions based on the physical insights obtained from StS analysis. It was concluded that while the centrifugal barrier-based approach captures dissociation/recombination kinetics well, it does not perform well in capturing the energy transfer dynamics. The VS binning strategy performs better than the conventional binning strategies due to the recombination phenomenon exhibiting a predominantly vibration-specific population distribution,especially at high vibrational quantum numbers. A hybrid binning strategy based on energy deficit from centrifugal barrier and inelastic kinetics outperforms all the binning strategies. 

\begin{acknowledgements}
    The authors would like to acknowledge funding support from the Air Force Office of Scientific Research (AFOSR) under AFOSR Grant No.: FA9550-22-1-0039.
\end{acknowledgements}

\bibliography{aipsamp}

\begin{thebibliography}{51}%
\makeatletter
\providecommand \@ifxundefined [1]{%
 \@ifx{#1\undefined}
}%
\providecommand \@ifnum [1]{%
 \ifnum #1\expandafter \@firstoftwo
 \else \expandafter \@secondoftwo
 \fi
}%
\providecommand \@ifx [1]{%
 \ifx #1\expandafter \@firstoftwo
 \else \expandafter \@secondoftwo
 \fi
}%
\providecommand \natexlab [1]{#1}%
\providecommand \enquote  [1]{``#1''}%
\providecommand \bibnamefont  [1]{#1}%
\providecommand \bibfnamefont [1]{#1}%
\providecommand \citenamefont [1]{#1}%
\providecommand \href@noop [0]{\@secondoftwo}%
\providecommand \href [0]{\begingroup \@sanitize@url \@href}%
\providecommand \@href[1]{\@@startlink{#1}\@@href}%
\providecommand \@@href[1]{\endgroup#1\@@endlink}%
\providecommand \@sanitize@url [0]{\catcode `\\12\catcode `\$12\catcode `\&12\catcode `\#12\catcode `\^12\catcode `\_12\catcode `\%12\relax}%
\providecommand \@@startlink[1]{}%
\providecommand \@@endlink[0]{}%
\providecommand \url  [0]{\begingroup\@sanitize@url \@url }%
\providecommand \@url [1]{\endgroup\@href {#1}{\urlprefix }}%
\providecommand \urlprefix  [0]{URL }%
\providecommand \Eprint [0]{\href }%
\providecommand \doibase [0]{http://dx.doi.org/}%
\providecommand \selectlanguage [0]{\@gobble}%
\providecommand \bibinfo  [0]{\@secondoftwo}%
\providecommand \bibfield  [0]{\@secondoftwo}%
\providecommand \translation [1]{[#1]}%
\providecommand \BibitemOpen [0]{}%
\providecommand \bibitemStop [0]{}%
\providecommand \bibitemNoStop [0]{.\EOS\space}%
\providecommand \EOS [0]{\spacefactor3000\relax}%
\providecommand \BibitemShut  [1]{\csname bibitem#1\endcsname}%
\let\auto@bib@innerbib\@empty
%</preamble>
\bibitem [{\citenamefont {Park}(1990)}]{Park}%
  \BibitemOpen
  \bibfield  {author} {\bibinfo {author} {\bibfnamefont {C.}~\bibnamefont {Park}},\ }\href@noop {} {\emph {\bibinfo {title} {Nonequilibrium Hypersonic Aerothermodynamics}}}\ (\bibinfo  {publisher} {John Wiley and Sons Inc},\ \bibinfo {year} {1990})\BibitemShut {NoStop}%
\bibitem [{\citenamefont {Gnoffo}, \citenamefont {Gupta},\ and\ \citenamefont {Shinn}(1989)}]{Gnoffo}%
  \BibitemOpen
  \bibfield  {author} {\bibinfo {author} {\bibfnamefont {P.~A.}\ \bibnamefont {Gnoffo}}, \bibinfo {author} {\bibfnamefont {R.~N.}\ \bibnamefont {Gupta}}, \ and\ \bibinfo {author} {\bibfnamefont {J.~L.}\ \bibnamefont {Shinn}},\ }\href@noop {} {\enquote {\bibinfo {title} {Conservation equations and physical models for hypersonic air flows in thermal and chemical nonequilibrium},}\ }\bibinfo {type} {Tech. Rep.}\ (\bibinfo  {institution} {NASA},\ \bibinfo {year} {1989})\BibitemShut {NoStop}%
\bibitem [{\citenamefont {Adamovich}\ \emph {et~al.}(1995)\citenamefont {Adamovich}, \citenamefont {Macheret}, \citenamefont {Rich},\ and\ \citenamefont {Treanor}}]{Adamovich}%
  \BibitemOpen
  \bibfield  {author} {\bibinfo {author} {\bibfnamefont {I.~V.}\ \bibnamefont {Adamovich}}, \bibinfo {author} {\bibfnamefont {S.~O.}\ \bibnamefont {Macheret}}, \bibinfo {author} {\bibfnamefont {J.~W.}\ \bibnamefont {Rich}}, \ and\ \bibinfo {author} {\bibfnamefont {C.~E.}\ \bibnamefont {Treanor}},\ }\bibfield  {title} {\enquote {\bibinfo {title} {Vibrational relaxation and dissociation behind shock waves part 1: Kinetic rate models},}\ }\href@noop {} {\bibfield  {journal} {\bibinfo  {journal} {AIAA}\ } (\bibinfo {year} {1995})}\BibitemShut {NoStop}%
\bibitem [{\citenamefont {Sharma}, \citenamefont {Huo},\ and\ \citenamefont {Park}(1992)}]{SSH_Approx}%
  \BibitemOpen
  \bibfield  {author} {\bibinfo {author} {\bibfnamefont {S.~P.}\ \bibnamefont {Sharma}}, \bibinfo {author} {\bibfnamefont {W.~M.}\ \bibnamefont {Huo}}, \ and\ \bibinfo {author} {\bibfnamefont {C.}~\bibnamefont {Park}},\ }\bibfield  {title} {\enquote {\bibinfo {title} {Rate parameters for coupled vibration-dissociation in a generalized ssh approximation},}\ }\href@noop {} {\bibfield  {journal} {\bibinfo  {journal} {Journal of Thermophysics and Heat Transfer}\ } (\bibinfo {year} {1992})}\BibitemShut {NoStop}%
\bibitem [{\citenamefont {Schwartz}, \citenamefont {Slawsky},\ and\ \citenamefont {Herzfeld}(1952)}]{Zener}%
  \BibitemOpen
  \bibfield  {author} {\bibinfo {author} {\bibfnamefont {R.}~\bibnamefont {Schwartz}}, \bibinfo {author} {\bibfnamefont {Z.}~\bibnamefont {Slawsky}}, \ and\ \bibinfo {author} {\bibfnamefont {K.}~\bibnamefont {Herzfeld}},\ }\bibfield  {title} {\enquote {\bibinfo {title} {Calculation of vibrational relaxation times in gases},}\ }\href@noop {} {\bibfield  {journal} {\bibinfo  {journal} {The Journal of Chemical Physics}\ } (\bibinfo {year} {1952})}\BibitemShut {NoStop}%
\bibitem [{\citenamefont {Park}(1987)}]{Parkmodel}%
  \BibitemOpen
  \bibfield  {author} {\bibinfo {author} {\bibfnamefont {C.}~\bibnamefont {Park}},\ }\bibfield  {title} {\enquote {\bibinfo {title} {Assessment of two-temperature kinetic model for ionizing air},}\ }\href@noop {} {\bibfield  {journal} {\bibinfo  {journal} {AIAA-87-1574}\ } (\bibinfo {year} {1987})}\BibitemShut {NoStop}%
\bibitem [{\citenamefont {Kim}\ and\ \citenamefont {Boyd}(2015)}]{Boyd_emp_sts}%
  \BibitemOpen
  \bibfield  {author} {\bibinfo {author} {\bibfnamefont {J.~G.}\ \bibnamefont {Kim}}\ and\ \bibinfo {author} {\bibfnamefont {I.~D.}\ \bibnamefont {Boyd}},\ }\bibfield  {title} {\enquote {\bibinfo {title} {Master equation analysis of post normal shock waves of nitrogen},}\ }\href@noop {} {\bibfield  {journal} {\bibinfo  {journal} {Journal of Thermophysics and Heat Transfer}\ } (\bibinfo {year} {2015})}\BibitemShut {NoStop}%
\bibitem [{\citenamefont {Kim}, \citenamefont {Kwon},\ and\ \citenamefont {Park}(2010)}]{Kim_Hydrogen}%
  \BibitemOpen
  \bibfield  {author} {\bibinfo {author} {\bibfnamefont {J.~G.}\ \bibnamefont {Kim}}, \bibinfo {author} {\bibfnamefont {O.~J.}\ \bibnamefont {Kwon}}, \ and\ \bibinfo {author} {\bibfnamefont {C.}~\bibnamefont {Park}},\ }\bibfield  {title} {\enquote {\bibinfo {title} {Master equation study and nonequilibrium chemical reactions for hydrogen molecule},}\ }\href@noop {} {\bibfield  {journal} {\bibinfo  {journal} {Journal of Thermophysics and Heat Transfer}\ } (\bibinfo {year} {2010})}\BibitemShut {NoStop}%
\bibitem [{\citenamefont {Panesi}\ \emph {et~al.}(2013)\citenamefont {Panesi}, \citenamefont {Jaffe}, \citenamefont {Schwenke},\ and\ \citenamefont {Magin}}]{Panesi2013_N3}%
  \BibitemOpen
  \bibfield  {author} {\bibinfo {author} {\bibfnamefont {M.}~\bibnamefont {Panesi}}, \bibinfo {author} {\bibfnamefont {R.~L.}\ \bibnamefont {Jaffe}}, \bibinfo {author} {\bibfnamefont {D.~W.}\ \bibnamefont {Schwenke}}, \ and\ \bibinfo {author} {\bibfnamefont {T.~E.}\ \bibnamefont {Magin}},\ }\bibfield  {title} {\enquote {\bibinfo {title} {Rovibrational internal energy transfer and dissociation of $\text{N}_2({}^1\sum_g^+)$-\text{N}(${}^4\text{S}_u$) system},}\ }\href {\doibase 10.1063/1.4774412} {\bibfield  {journal} {\bibinfo  {journal} {Journal of Chemical Physics}\ }\textbf {\bibinfo {volume} {138}} (\bibinfo {year} {2013}),\ 10.1063/1.4774412}\BibitemShut {NoStop}%
\bibitem [{\citenamefont {Jo}\ \emph {et~al.}(2022)\citenamefont {Jo}, \citenamefont {Venturi}, \citenamefont {Sharma}, \citenamefont {Munafò},\ and\ \citenamefont {Panesi}}]{Sung_Min_N2O}%
  \BibitemOpen
  \bibfield  {author} {\bibinfo {author} {\bibfnamefont {S.~M.}\ \bibnamefont {Jo}}, \bibinfo {author} {\bibfnamefont {S.}~\bibnamefont {Venturi}}, \bibinfo {author} {\bibfnamefont {M.~P.}\ \bibnamefont {Sharma}}, \bibinfo {author} {\bibfnamefont {A.}~\bibnamefont {Munafò}}, \ and\ \bibinfo {author} {\bibfnamefont {M.}~\bibnamefont {Panesi}},\ }\bibfield  {title} {\enquote {\bibinfo {title} {Rovibrational-specific qct and master equation study on $\text{N}_2(\text{X}^1 \sum^+_g) + \text{O}({}^3\text{P})$ and $\text{NO}(\text{X}^2 \prod) + \text{N}({}^4\text{S})$ systems in high-energy collisions},}\ }\href {\doibase 10.1021/acs.jpca.1c10346} {\bibfield  {journal} {\bibinfo  {journal} {The Journal of Physical Chemistry A}\ }\textbf {\bibinfo {volume} {126}},\ \bibinfo {pages} {3273--3290} (\bibinfo {year} {2022})},\ \bibinfo {note} {pMID: 35604650},\ \Eprint {http://arxiv.org/abs/https://doi.org/10.1021/acs.jpca.1c10346} {https://doi.org/10.1021/acs.jpca.1c10346} \BibitemShut {NoStop}%
\bibitem [{\citenamefont {Kim}\ and\ \citenamefont {Boyd}(2013)}]{Boyd_Sts_N2}%
  \BibitemOpen
  \bibfield  {author} {\bibinfo {author} {\bibfnamefont {J.~G.}\ \bibnamefont {Kim}}\ and\ \bibinfo {author} {\bibfnamefont {I.~D.}\ \bibnamefont {Boyd}},\ }\bibfield  {title} {\enquote {\bibinfo {title} {State-resolved master equation analysis of thermochemical non-equilibrium of nitrogen},}\ }\href@noop {} {\bibfield  {journal} {\bibinfo  {journal} {Chemical Physics}\ } (\bibinfo {year} {2013})}\BibitemShut {NoStop}%
\bibitem [{\citenamefont {Andrienko}\ and\ \citenamefont {Boyd}(2016)}]{Boyd_O2}%
  \BibitemOpen
  \bibfield  {author} {\bibinfo {author} {\bibfnamefont {D.~A.}\ \bibnamefont {Andrienko}}\ and\ \bibinfo {author} {\bibfnamefont {I.~D.}\ \bibnamefont {Boyd}},\ }\bibfield  {title} {\enquote {\bibinfo {title} {Rovibrational energy transfer and dissociation in $\text{O}_2+\text{O}$ collisions},}\ }\href@noop {} {\bibfield  {journal} {\bibinfo  {journal} {The Journal of Chemical Physics}\ } (\bibinfo {year} {2016})}\BibitemShut {NoStop}%
\bibitem [{\citenamefont {Macdonald}\ \emph {et~al.}(2020)\citenamefont {Macdonald}, \citenamefont {Torres}, \citenamefont {Schwartzentruber},\ and\ \citenamefont {Panesi}}]{Macdonald2020_N3}%
  \BibitemOpen
  \bibfield  {author} {\bibinfo {author} {\bibfnamefont {R.~L.}\ \bibnamefont {Macdonald}}, \bibinfo {author} {\bibfnamefont {E.}~\bibnamefont {Torres}}, \bibinfo {author} {\bibfnamefont {T.~E.}\ \bibnamefont {Schwartzentruber}}, \ and\ \bibinfo {author} {\bibfnamefont {M.}~\bibnamefont {Panesi}},\ }\bibfield  {title} {\enquote {\bibinfo {title} {State-to-state master equation and direct molecular simulation study of energy transfer and dissociation for the $\text{N}_2$-\text{N} system},}\ }\href {\doibase 10.1021/acs.jpca.0c04029} {\bibfield  {journal} {\bibinfo  {journal} {Journal of Physical Chemistry A}\ }\textbf {\bibinfo {volume} {124}},\ \bibinfo {pages} {6986--7000} (\bibinfo {year} {2020})}\BibitemShut {NoStop}%
\bibitem [{\citenamefont {Venturi}\ \emph {et~al.}(2020)\citenamefont {Venturi}, \citenamefont {Sharma}, \citenamefont {Lopez},\ and\ \citenamefont {Panesi}}]{Venturi2020_O3Diss}%
  \BibitemOpen
  \bibfield  {author} {\bibinfo {author} {\bibfnamefont {S.}~\bibnamefont {Venturi}}, \bibinfo {author} {\bibfnamefont {M.~P.}\ \bibnamefont {Sharma}}, \bibinfo {author} {\bibfnamefont {B.}~\bibnamefont {Lopez}}, \ and\ \bibinfo {author} {\bibfnamefont {M.}~\bibnamefont {Panesi}},\ }\bibfield  {title} {\enquote {\bibinfo {title} {Data-inspired and physics-driven model reduction for dissociation: Application to the $\text{O}_2$+\text{O} system},}\ }\href {\doibase 10.1021/acs.jpca.0c04516} {\bibfield  {journal} {\bibinfo  {journal} {Journal of Physical Chemistry A}\ }\textbf {\bibinfo {volume} {124}},\ \bibinfo {pages} {8359--8372} (\bibinfo {year} {2020})}\BibitemShut {NoStop}%
\bibitem [{\citenamefont {Munaf\`{o}}\ \emph {et~al.}(2020)\citenamefont {Munaf\`{o}}, \citenamefont {Venturi}, \citenamefont {Sharma},\ and\ \citenamefont {Panesi}}]{Munafo2020_N2O}%
  \BibitemOpen
  \bibfield  {author} {\bibinfo {author} {\bibfnamefont {A.}~\bibnamefont {Munaf\`{o}}}, \bibinfo {author} {\bibfnamefont {S.}~\bibnamefont {Venturi}}, \bibinfo {author} {\bibfnamefont {M.~P.}\ \bibnamefont {Sharma}}, \ and\ \bibinfo {author} {\bibfnamefont {M.}~\bibnamefont {Panesi}},\ }\bibfield  {title} {\enquote {\bibinfo {title} {{Reduced-Order Modeling for Non-Equilibrium Air Flows}},}\ }in\ \href@noop {} {\emph {\bibinfo {booktitle} {AIAA Scitech 2020 Forum, AIAA 2020-1226}}}\ (\bibinfo {year} {2020})\BibitemShut {NoStop}%
\bibitem [{\citenamefont {Sahai}\ \emph {et~al.}(2017)\citenamefont {Sahai}, \citenamefont {Lopez}, \citenamefont {Johnston},\ and\ \citenamefont {Panesi}}]{Sahai2017_Adaptive}%
  \BibitemOpen
  \bibfield  {author} {\bibinfo {author} {\bibfnamefont {A.}~\bibnamefont {Sahai}}, \bibinfo {author} {\bibfnamefont {B.}~\bibnamefont {Lopez}}, \bibinfo {author} {\bibfnamefont {C.~O.}\ \bibnamefont {Johnston}}, \ and\ \bibinfo {author} {\bibfnamefont {M.}~\bibnamefont {Panesi}},\ }\bibfield  {title} {\enquote {\bibinfo {title} {Adaptive coarse graining method for energy transfer and dissociation kinetics of polyatomic species},}\ }\href {\doibase 10.1063/1.4996654} {\bibfield  {journal} {\bibinfo  {journal} {Journal of Chemical Physics}\ }\textbf {\bibinfo {volume} {147}} (\bibinfo {year} {2017}),\ 10.1063/1.4996654}\BibitemShut {NoStop}%
\bibitem [{\citenamefont {Macdonald}\ \emph {et~al.}(2018{\natexlab{a}})\citenamefont {Macdonald}, \citenamefont {Jaffe}, \citenamefont {Schwenke},\ and\ \citenamefont {Panesi}}]{Macdonald2018_CGQCT}%
  \BibitemOpen
  \bibfield  {author} {\bibinfo {author} {\bibfnamefont {R.~L.}\ \bibnamefont {Macdonald}}, \bibinfo {author} {\bibfnamefont {R.~L.}\ \bibnamefont {Jaffe}}, \bibinfo {author} {\bibfnamefont {D.~W.}\ \bibnamefont {Schwenke}}, \ and\ \bibinfo {author} {\bibfnamefont {M.}~\bibnamefont {Panesi}},\ }\bibfield  {title} {\enquote {\bibinfo {title} {Construction of a coarse-grain quasi-classical trajectory method. i. theory and application to $\text{N}_2$–$\text{N}_2$ system},}\ }\href {\doibase 10.1063/1.5011331} {\bibfield  {journal} {\bibinfo  {journal} {Journal of Chemical Physics}\ }\textbf {\bibinfo {volume} {148}} (\bibinfo {year} {2018}{\natexlab{a}}),\ 10.1063/1.5011331}\BibitemShut {NoStop}%
\bibitem [{\citenamefont {Macdonald}\ \emph {et~al.}(2018{\natexlab{b}})\citenamefont {Macdonald}, \citenamefont {Grover}, \citenamefont {Schwartzentruber},\ and\ \citenamefont {Panesi}}]{Macdonald2018_CGQCT2}%
  \BibitemOpen
  \bibfield  {author} {\bibinfo {author} {\bibfnamefont {R.~L.}\ \bibnamefont {Macdonald}}, \bibinfo {author} {\bibfnamefont {M.}~\bibnamefont {Grover}}, \bibinfo {author} {\bibfnamefont {T.}~\bibnamefont {Schwartzentruber}}, \ and\ \bibinfo {author} {\bibfnamefont {M.}~\bibnamefont {Panesi}},\ }\bibfield  {title} {\enquote {\bibinfo {title} {Construction of a coarse-grain quasi-classical trajectory method. ii. comparison against the direct molecular simulation method},}\ }\href {\doibase 10.1063/1.5011332} {\bibfield  {journal} {\bibinfo  {journal} {Journal of Chemical Physics}\ }\textbf {\bibinfo {volume} {148}} (\bibinfo {year} {2018}{\natexlab{b}}),\ 10.1063/1.5011332}\BibitemShut {NoStop}%
\bibitem [{\citenamefont {Macdonald}, \citenamefont {Jaffe},\ and\ \citenamefont {Panesi}(2019)}]{Robyn_N3_CGQCT}%
  \BibitemOpen
  \bibfield  {author} {\bibinfo {author} {\bibfnamefont {R.~L.}\ \bibnamefont {Macdonald}}, \bibinfo {author} {\bibfnamefont {R.~L.}\ \bibnamefont {Jaffe}}, \ and\ \bibinfo {author} {\bibfnamefont {M.}~\bibnamefont {Panesi}},\ }\bibfield  {title} {\enquote {\bibinfo {title} {Hybrid reduced order model for $\text{N}_2$-n interactions for application to dissociation and energy tranfer processes},}\ }\href@noop {} {\bibfield  {journal} {\bibinfo  {journal} {AIP Conference Proceedings}\ } (\bibinfo {year} {2019})}\BibitemShut {NoStop}%
\bibitem [{\citenamefont {Colonna}, \citenamefont {Pietanza},\ and\ \citenamefont {Capitelli}(2011)}]{Colonna_TLD_2011}%
  \BibitemOpen
  \bibfield  {author} {\bibinfo {author} {\bibfnamefont {G.}~\bibnamefont {Colonna}}, \bibinfo {author} {\bibfnamefont {L.~D.}\ \bibnamefont {Pietanza}}, \ and\ \bibinfo {author} {\bibfnamefont {M.}~\bibnamefont {Capitelli}},\ }\bibfield  {title} {\enquote {\bibinfo {title} {Reduced two-level approach for air kinetics in recombination regime},}\ }\href@noop {} {\bibfield  {journal} {\bibinfo  {journal} {27th International Symposium on Rarefied Gas Dynamics}\ } (\bibinfo {year} {2011})}\BibitemShut {NoStop}%
\bibitem [{\citenamefont {Colonna}, \citenamefont {Pietanza},\ and\ \citenamefont {Capitelli}(2012)}]{Colonna_TLD_2012}%
  \BibitemOpen
  \bibfield  {author} {\bibinfo {author} {\bibfnamefont {G.}~\bibnamefont {Colonna}}, \bibinfo {author} {\bibfnamefont {L.~D.}\ \bibnamefont {Pietanza}}, \ and\ \bibinfo {author} {\bibfnamefont {M.}~\bibnamefont {Capitelli}},\ }\bibfield  {title} {\enquote {\bibinfo {title} {Macroscopic kinetic model for air in nozzle flow},}\ }\href@noop {} {\bibfield  {journal} {\bibinfo  {journal} {28th International Symposium on Rarefied Gas Dynamics}\ } (\bibinfo {year} {2012})}\BibitemShut {NoStop}%
\bibitem [{\citenamefont {Zanardi}\ \emph {et~al.}(2025{\natexlab{a}})\citenamefont {Zanardi}, \citenamefont {Padovan}, \citenamefont {Bodony},\ and\ \citenamefont {Panesi}}]{Zanardi_PG}%
  \BibitemOpen
  \bibfield  {author} {\bibinfo {author} {\bibfnamefont {I.}~\bibnamefont {Zanardi}}, \bibinfo {author} {\bibfnamefont {A.}~\bibnamefont {Padovan}}, \bibinfo {author} {\bibfnamefont {D.}~\bibnamefont {Bodony}}, \ and\ \bibinfo {author} {\bibfnamefont {M.}~\bibnamefont {Panesi}},\ }\bibfield  {title} {\enquote {\bibinfo {title} {Petrov-galerkin model reduction for thermochemical nonequilibrium gas mixtures},}\ }\href@noop {} {\bibfield  {journal} {\bibinfo  {journal} {Journal of Computational Physics}\ } (\bibinfo {year} {2025}{\natexlab{a}})}\BibitemShut {NoStop}%
\bibitem [{\citenamefont {Zanardi}\ \emph {et~al.}(2025{\natexlab{b}})\citenamefont {Zanardi}, \citenamefont {Meini}, \citenamefont {Padovan}, \citenamefont {Bodony},\ and\ \citenamefont {Panesi}}]{Zanardi_Ar}%
  \BibitemOpen
  \bibfield  {author} {\bibinfo {author} {\bibfnamefont {I.}~\bibnamefont {Zanardi}}, \bibinfo {author} {\bibfnamefont {A.}~\bibnamefont {Meini}}, \bibinfo {author} {\bibfnamefont {A.}~\bibnamefont {Padovan}}, \bibinfo {author} {\bibfnamefont {D.}~\bibnamefont {Bodony}}, \ and\ \bibinfo {author} {\bibfnamefont {M.}~\bibnamefont {Panesi}},\ }\bibfield  {title} {\enquote {\bibinfo {title} {Petrov-galerkin model reduction for collisional-radiative argon plasma},}\ }\href@noop {} {\bibfield  {journal} {\bibinfo  {journal} {arXiv preprint arXiv:2506.05483}\ } (\bibinfo {year} {2025}{\natexlab{b}})}\BibitemShut {NoStop}%
\bibitem [{\citenamefont {Liu}\ \emph {et~al.}(2015)\citenamefont {Liu}, \citenamefont {Panesi}, \citenamefont {Sahai},\ and\ \citenamefont {Vinokur}}]{Liu2015}%
  \BibitemOpen
  \bibfield  {author} {\bibinfo {author} {\bibfnamefont {Y.}~\bibnamefont {Liu}}, \bibinfo {author} {\bibfnamefont {M.}~\bibnamefont {Panesi}}, \bibinfo {author} {\bibfnamefont {A.}~\bibnamefont {Sahai}}, \ and\ \bibinfo {author} {\bibfnamefont {M.}~\bibnamefont {Vinokur}},\ }\bibfield  {title} {\enquote {\bibinfo {title} {General multi-group macroscopic modeling for thermo-chemical non-equilibrium gas mixtures},}\ }\href {\doibase 10.1063/1.4915926} {\bibfield  {journal} {\bibinfo  {journal} {Journal of Chemical Physics}\ }\textbf {\bibinfo {volume} {142}} (\bibinfo {year} {2015}),\ 10.1063/1.4915926}\BibitemShut {NoStop}%
\bibitem [{\citenamefont {Neitzel}, \citenamefont {Andrienko},\ and\ \citenamefont {Boyd}(2017)}]{Boyd_O21}%
  \BibitemOpen
  \bibfield  {author} {\bibinfo {author} {\bibfnamefont {K.}~\bibnamefont {Neitzel}}, \bibinfo {author} {\bibfnamefont {D.}~\bibnamefont {Andrienko}}, \ and\ \bibinfo {author} {\bibfnamefont {I.~D.}\ \bibnamefont {Boyd}},\ }\bibfield  {title} {\enquote {\bibinfo {title} {Aerothermochemical nonequilibrium modeling for oxygen flows},}\ }\href@noop {} {\bibfield  {journal} {\bibinfo  {journal} {Journal of Thermophysics and Heat Transfer}\ } (\bibinfo {year} {2017})}\BibitemShut {NoStop}%
\bibitem [{\citenamefont {Munafo\`''}\ \emph {et~al.}(2016)\citenamefont {Munafo\`''}, \citenamefont {Venturi}, \citenamefont {Macdonald},\ and\ \citenamefont {Panesi}}]{Munafo_recomb}%
  \BibitemOpen
  \bibfield  {author} {\bibinfo {author} {\bibfnamefont {A.}~\bibnamefont {Munafo\`''}}, \bibinfo {author} {\bibfnamefont {S.}~\bibnamefont {Venturi}}, \bibinfo {author} {\bibfnamefont {R.}~\bibnamefont {Macdonald}}, \ and\ \bibinfo {author} {\bibfnamefont {M.}~\bibnamefont {Panesi}},\ }\bibfield  {title} {\enquote {\bibinfo {title} {State-to-state and reduced-order models for recombination and energy transfer in aerothermal environments},}\ }\href@noop {} {\bibfield  {journal} {\bibinfo  {journal} {AIAA SciTech Forum}\ } (\bibinfo {year} {2016})}\BibitemShut {NoStop}%
\bibitem [{\citenamefont {Capitelli}, \citenamefont {Armenise},\ and\ \citenamefont {Gorse}(1997)}]{Capitelli_BL}%
  \BibitemOpen
  \bibfield  {author} {\bibinfo {author} {\bibfnamefont {M.}~\bibnamefont {Capitelli}}, \bibinfo {author} {\bibfnamefont {I.}~\bibnamefont {Armenise}}, \ and\ \bibinfo {author} {\bibfnamefont {C.}~\bibnamefont {Gorse}},\ }\bibfield  {title} {\enquote {\bibinfo {title} {State-to-state approach in the kinetics of air components under re-entry conditions},}\ }\href@noop {} {\bibfield  {journal} {\bibinfo  {journal} {Journal of Thermophysics and Heat Transfer}\ } (\bibinfo {year} {1997})}\BibitemShut {NoStop}%
\bibitem [{\citenamefont {Colonna}\ \emph {et~al.}(1999)\citenamefont {Colonna}, \citenamefont {Tuttafesta}, \citenamefont {Capitelli},\ and\ \citenamefont {Giordano}}]{Colonna_1999}%
  \BibitemOpen
  \bibfield  {author} {\bibinfo {author} {\bibfnamefont {G.}~\bibnamefont {Colonna}}, \bibinfo {author} {\bibfnamefont {M.}~\bibnamefont {Tuttafesta}}, \bibinfo {author} {\bibfnamefont {M.}~\bibnamefont {Capitelli}}, \ and\ \bibinfo {author} {\bibfnamefont {D.}~\bibnamefont {Giordano}},\ }\bibfield  {title} {\enquote {\bibinfo {title} {Non-{A}rrhenius $\rm{NO}$ formation rate in one-dimensional nozzle airflow},}\ }\href@noop {} {\bibfield  {journal} {\bibinfo  {journal} {J. Thermophys. Heat Transfer}\ }\textbf {\bibinfo {volume} {13}},\ \bibinfo {pages} {372--375} (\bibinfo {year} {1999})}\BibitemShut {NoStop}%
\bibitem [{\citenamefont {Colonna}\ and\ \citenamefont {Capitelli}(2001)}]{Colonna_2001}%
  \BibitemOpen
  \bibfield  {author} {\bibinfo {author} {\bibfnamefont {G.}~\bibnamefont {Colonna}}\ and\ \bibinfo {author} {\bibfnamefont {M.}~\bibnamefont {Capitelli}},\ }\bibfield  {title} {\enquote {\bibinfo {title} {Self-consistent model of chemical, vibrational, electron kinetics in nozzle expansion},}\ }\href@noop {} {\bibfield  {journal} {\bibinfo  {journal} {J. Thermophys. Heat Transfer}\ }\textbf {\bibinfo {volume} {15}},\ \bibinfo {pages} {308--316} (\bibinfo {year} {2001})}\BibitemShut {NoStop}%
\bibitem [{\citenamefont {Pan}, \citenamefont {Wilson},\ and\ \citenamefont {Stephani}(2019)}]{VibStS_Recomb_DSMC}%
  \BibitemOpen
  \bibfield  {author} {\bibinfo {author} {\bibfnamefont {T.-J.}\ \bibnamefont {Pan}}, \bibinfo {author} {\bibfnamefont {T.~J.}\ \bibnamefont {Wilson}}, \ and\ \bibinfo {author} {\bibfnamefont {K.~A.}\ \bibnamefont {Stephani}},\ }\bibfield  {title} {\enquote {\bibinfo {title} {Vibrational state-specific model for dissociation and recombination of the $o_{2}(^3\sum^{+}_{g})+o(^{3}p_{2})$ system in dsmc},}\ }\href@noop {} {\bibfield  {journal} {\bibinfo  {journal} {The Journal of Chemical Physics}\ } (\bibinfo {year} {2019})}\BibitemShut {NoStop}%
\bibitem [{\citenamefont {Macdonald}(2024)}]{Robyn_recomb}%
  \BibitemOpen
  \bibfield  {author} {\bibinfo {author} {\bibfnamefont {R.~L.}\ \bibnamefont {Macdonald}},\ }\bibfield  {title} {\enquote {\bibinfo {title} {State-to-state study of non-equilibrium recombination of oxygen and nitrogen molecules},}\ }\href@noop {} {\bibfield  {journal} {\bibinfo  {journal} {The Journal of Chemical Physics}\ } (\bibinfo {year} {2024})}\BibitemShut {NoStop}%
\bibitem [{\citenamefont {Esposito}\ \emph {et~al.}(2008)\citenamefont {Esposito}, \citenamefont {Armenise}, \citenamefont {Capitta},\ and\ \citenamefont {Capitelli}}]{Capitelli_O3_QCT}%
  \BibitemOpen
  \bibfield  {author} {\bibinfo {author} {\bibfnamefont {F.}~\bibnamefont {Esposito}}, \bibinfo {author} {\bibfnamefont {I.}~\bibnamefont {Armenise}}, \bibinfo {author} {\bibfnamefont {G.}~\bibnamefont {Capitta}}, \ and\ \bibinfo {author} {\bibfnamefont {M.}~\bibnamefont {Capitelli}},\ }\bibfield  {title} {\enquote {\bibinfo {title} {O-$\text{O}_2$ state-to-state vibrational relaxation and dissociation rates based on quasiclassical calculations},}\ }\href@noop {} {\bibfield  {journal} {\bibinfo  {journal} {Chemical Physics}\ } (\bibinfo {year} {2008})}\BibitemShut {NoStop}%
\bibitem [{\citenamefont {Chaudhry}\ \emph {et~al.}(2018)\citenamefont {Chaudhry}, \citenamefont {Grover}, \citenamefont {Bender}, \citenamefont {Schwartzentruber},\ and\ \citenamefont {Candler}}]{Bender_QCT}%
  \BibitemOpen
  \bibfield  {author} {\bibinfo {author} {\bibfnamefont {R.~S.}\ \bibnamefont {Chaudhry}}, \bibinfo {author} {\bibfnamefont {M.~S.}\ \bibnamefont {Grover}}, \bibinfo {author} {\bibfnamefont {J.~D.}\ \bibnamefont {Bender}}, \bibinfo {author} {\bibfnamefont {T.~E.}\ \bibnamefont {Schwartzentruber}}, \ and\ \bibinfo {author} {\bibfnamefont {G.~V.}\ \bibnamefont {Candler}},\ }\bibfield  {title} {\enquote {\bibinfo {title} {Quasiclassical trajectory analysis of oxygen dissociation via $\text{O}_2$, o, and $\text{N}_2$},}\ }\href@noop {} {\bibfield  {journal} {\bibinfo  {journal} {AIAA SciTech Forum}\ } (\bibinfo {year} {January, 2018})}\BibitemShut {NoStop}%
\bibitem [{\citenamefont {Venturi}, \citenamefont {Jaffe},\ and\ \citenamefont {Panesi}(2020)}]{Venturi2020_O3Bayesian}%
  \BibitemOpen
  \bibfield  {author} {\bibinfo {author} {\bibfnamefont {S.}~\bibnamefont {Venturi}}, \bibinfo {author} {\bibfnamefont {R.~L.}\ \bibnamefont {Jaffe}}, \ and\ \bibinfo {author} {\bibfnamefont {M.}~\bibnamefont {Panesi}},\ }\bibfield  {title} {\enquote {\bibinfo {title} {Bayesian machine learning approach to the quantification of uncertainties on ab initio potential energy surfaces},}\ }\href {\doibase 10.1021/acs.jpca.0c02395} {\bibfield  {journal} {\bibinfo  {journal} {Journal of Physical Chemistry A}\ }\textbf {\bibinfo {volume} {124}},\ \bibinfo {pages} {5129--5146} (\bibinfo {year} {2020})}\BibitemShut {NoStop}%
\bibitem [{\citenamefont {Schwenke}(1988)}]{Schwenke1988_VVTC}%
  \BibitemOpen
  \bibfield  {author} {\bibinfo {author} {\bibfnamefont {D.~W.}\ \bibnamefont {Schwenke}},\ }\bibfield  {title} {\enquote {\bibinfo {title} {Calculations of rate constants for the three-body recombination of $\text{H}_2$ in the presence of $\text{H}_2$},}\ }\href {\doibase 10.1063/1.455104} {\bibfield  {journal} {\bibinfo  {journal} {Journal of Chemical Physics}\ }\textbf {\bibinfo {volume} {89}},\ \bibinfo {pages} {2076--2091} (\bibinfo {year} {1988})}\BibitemShut {NoStop}%
\bibitem [{\citenamefont {Jaffe}, \citenamefont {Schwenke},\ and\ \citenamefont {Chaban}(2008)}]{Jaffe_N3_PES1}%
  \BibitemOpen
  \bibfield  {author} {\bibinfo {author} {\bibfnamefont {R.}~\bibnamefont {Jaffe}}, \bibinfo {author} {\bibfnamefont {D.~W.}\ \bibnamefont {Schwenke}}, \ and\ \bibinfo {author} {\bibfnamefont {G.}~\bibnamefont {Chaban}},\ }\bibfield  {title} {\enquote {\bibinfo {title} {Vibrational and rotational excitation and relaxation of nitrogen from accurate theoretical calculations},}\ }\href@noop {} {\bibfield  {journal} {\bibinfo  {journal} {46th AIAA Aerospace Sciences Meeting and Exhibit}\ } (\bibinfo {year} {2008})}\BibitemShut {NoStop}%
\bibitem [{\citenamefont {Chaban}\ \emph {et~al.}(2008)\citenamefont {Chaban}, \citenamefont {Jaffe}, \citenamefont {Schwenke},\ and\ \citenamefont {Huo}}]{Chaban_N3_PES2}%
  \BibitemOpen
  \bibfield  {author} {\bibinfo {author} {\bibfnamefont {G.}~\bibnamefont {Chaban}}, \bibinfo {author} {\bibfnamefont {R.}~\bibnamefont {Jaffe}}, \bibinfo {author} {\bibfnamefont {D.}~\bibnamefont {Schwenke}}, \ and\ \bibinfo {author} {\bibfnamefont {W.}~\bibnamefont {Huo}},\ }\bibfield  {title} {\enquote {\bibinfo {title} {Dissociation cross sections and rate coefficients for nitrogen from accurate theoretical calculations},}\ }\href@noop {} {\bibfield  {journal} {\bibinfo  {journal} {46th AIAA Aerospace Sciences Meeting and Exhibit}\ } (\bibinfo {year} {2008})}\BibitemShut {NoStop}%
\bibitem [{\citenamefont {Varga}, \citenamefont {Paukku},\ and\ \citenamefont {Truhlar}(2017)}]{Varga_O3_PES}%
  \BibitemOpen
  \bibfield  {author} {\bibinfo {author} {\bibfnamefont {Z.}~\bibnamefont {Varga}}, \bibinfo {author} {\bibfnamefont {Y.}~\bibnamefont {Paukku}}, \ and\ \bibinfo {author} {\bibfnamefont {D.~G.}\ \bibnamefont {Truhlar}},\ }\bibfield  {title} {\enquote {\bibinfo {title} {Potential energy surfaces for o+o2 collisions},}\ }\href@noop {} {\bibfield  {journal} {\bibinfo  {journal} {The Journal of Chemical Physics}\ } (\bibinfo {year} {2017})}\BibitemShut {NoStop}%
\bibitem [{\citenamefont {Hindmarsh}\ \emph {et~al.}(2005)\citenamefont {Hindmarsh}, \citenamefont {Brown}, \citenamefont {Grant}, \citenamefont {Lee}, \citenamefont {Serban}, \citenamefont {Shumaker},\ and\ \citenamefont {Woodward}}]{hindmarsh2005sundials}%
  \BibitemOpen
  \bibfield  {author} {\bibinfo {author} {\bibfnamefont {A.~C.}\ \bibnamefont {Hindmarsh}}, \bibinfo {author} {\bibfnamefont {P.~N.}\ \bibnamefont {Brown}}, \bibinfo {author} {\bibfnamefont {K.~E.}\ \bibnamefont {Grant}}, \bibinfo {author} {\bibfnamefont {S.~L.}\ \bibnamefont {Lee}}, \bibinfo {author} {\bibfnamefont {R.}~\bibnamefont {Serban}}, \bibinfo {author} {\bibfnamefont {D.~E.}\ \bibnamefont {Shumaker}}, \ and\ \bibinfo {author} {\bibfnamefont {C.~S.}\ \bibnamefont {Woodward}},\ }\bibfield  {title} {\enquote {\bibinfo {title} {{SUNDIALS}: Suite of nonlinear and differential/algebraic equation solvers},}\ }\href@noop {} {\bibfield  {journal} {\bibinfo  {journal} {ACM Transactions on Mathematical Software}\ }\textbf {\bibinfo {volume} {31}},\ \bibinfo {pages} {363--396} (\bibinfo {year} {2005})}\BibitemShut {NoStop}%
\bibitem [{\citenamefont {Munafò}\ and\ \citenamefont {Panesi}(2023)}]{PLATO2023}%
  \BibitemOpen
  \bibfield  {author} {\bibinfo {author} {\bibfnamefont {A.}~\bibnamefont {Munafò}}\ and\ \bibinfo {author} {\bibfnamefont {M.}~\bibnamefont {Panesi}},\ }\bibfield  {title} {\enquote {\bibinfo {title} {Plato: a high-fidelity tool for multi-component plasmas},}\ }\href@noop {} {\bibfield  {journal} {\bibinfo  {journal} {AIAA 2023 - AIAA Aviation Forum, San Diego, CA}\ } (\bibinfo {year} {2023})}\BibitemShut {NoStop}%
\bibitem [{\citenamefont {Munaf\`{o}}\ and\ \citenamefont {Panesi}(2025)}]{Munafo_JTHT_plato_2025}%
  \BibitemOpen
  \bibfield  {author} {\bibinfo {author} {\bibfnamefont {A.}~\bibnamefont {Munaf\`{o}}}\ and\ \bibinfo {author} {\bibfnamefont {M.}~\bibnamefont {Panesi}},\ }\bibfield  {title} {\enquote {\bibinfo {title} {Plato: a high-fidelity library for multicomponent gases and plasmas},}\ }\href@noop {} {\bibfield  {journal} {\bibinfo  {journal} {J. Thermophys. Heat Transfer}\ } (\bibinfo {year} {2025})},\ \bibinfo {note} {in press}\BibitemShut {NoStop}%
\bibitem [{\citenamefont {Bourdon}\ and\ \citenamefont {Vervisch}(1996)}]{Bourdon}%
  \BibitemOpen
  \bibfield  {author} {\bibinfo {author} {\bibfnamefont {A.}~\bibnamefont {Bourdon}}\ and\ \bibinfo {author} {\bibfnamefont {P.}~\bibnamefont {Vervisch}},\ }\bibfield  {title} {\enquote {\bibinfo {title} {Three-body recombination rate of atomic nitrogen in low-pressure plasma flows},}\ }\href@noop {} {\bibfield  {journal} {\bibinfo  {journal} {Physical Review E}\ } (\bibinfo {year} {1996})}\BibitemShut {NoStop}%
\bibitem [{\citenamefont {Kim}\ and\ \citenamefont {Jo}(2021)}]{Kim_Jo}%
  \BibitemOpen
  \bibfield  {author} {\bibinfo {author} {\bibfnamefont {J.~G.}\ \bibnamefont {Kim}}\ and\ \bibinfo {author} {\bibfnamefont {S.~M.}\ \bibnamefont {Jo}},\ }\bibfield  {title} {\enquote {\bibinfo {title} {Modification of chemical-kinetic parameters for 11-air species in re-entry flows},}\ }\href@noop {} {\bibfield  {journal} {\bibinfo  {journal} {International Journal of Heat and Mass Transfer}\ } (\bibinfo {year} {2021})}\BibitemShut {NoStop}%
\bibitem [{\citenamefont {Hurle}(1971)}]{Tau_Expanding1}%
  \BibitemOpen
  \bibfield  {author} {\bibinfo {author} {\bibfnamefont {I.}~\bibnamefont {Hurle}},\ }\bibfield  {title} {\enquote {\bibinfo {title} {Nonequilibrium flows with special reference to the nozzle-flow problem},}\ }\href@noop {} {\bibfield  {journal} {\bibinfo  {journal} {Proceedings of the 8th International Shock Tube Symposium}\ } (\bibinfo {year} {1971})}\BibitemShut {NoStop}%
\bibitem [{\citenamefont {Hurle}, \citenamefont {Russo},\ and\ \citenamefont {Hall}(1964)}]{Tau_Expanding2}%
  \BibitemOpen
  \bibfield  {author} {\bibinfo {author} {\bibfnamefont {I.}~\bibnamefont {Hurle}}, \bibinfo {author} {\bibfnamefont {A.}~\bibnamefont {Russo}}, \ and\ \bibinfo {author} {\bibfnamefont {J.}~\bibnamefont {Hall}},\ }\bibfield  {title} {\enquote {\bibinfo {title} {Spectroscopic studies of vibrational nonequilibrium in supersonic nozzle flows},}\ }\href@noop {} {\bibfield  {journal} {\bibinfo  {journal} {Journal of Chemical Physics, Vol. 40, No.8}\ } (\bibinfo {year} {1964})}\BibitemShut {NoStop}%
\bibitem [{\citenamefont {MacDermott}\ and\ \citenamefont {Marshall}(1969)}]{Tau_Expanding3}%
  \BibitemOpen
  \bibfield  {author} {\bibinfo {author} {\bibfnamefont {W.}~\bibnamefont {MacDermott}}\ and\ \bibinfo {author} {\bibfnamefont {J.}~\bibnamefont {Marshall}},\ }\bibfield  {title} {\enquote {\bibinfo {title} {Nonequilibrium nozzle expansions of partially dissociated air: A comparison of theory and electron-beam experiments},}\ }\href@noop {} {\bibfield  {journal} {\bibinfo  {journal} {Arnold Engineering Development Center, AEDC-TR-69-66}\ } (\bibinfo {year} {1969})}\BibitemShut {NoStop}%
\bibitem [{\citenamefont {Howard}\ \emph {et~al.}(1990)\citenamefont {Howard}, \citenamefont {Dietz}, \citenamefont {McGregor},\ and\ \citenamefont {Limbaugh}}]{Tau_Expanding4}%
  \BibitemOpen
  \bibfield  {author} {\bibinfo {author} {\bibfnamefont {R.}~\bibnamefont {Howard}}, \bibinfo {author} {\bibfnamefont {K.}~\bibnamefont {Dietz}}, \bibinfo {author} {\bibfnamefont {W.}~\bibnamefont {McGregor}}, \ and\ \bibinfo {author} {\bibfnamefont {C.}~\bibnamefont {Limbaugh}},\ }\bibfield  {title} {\enquote {\bibinfo {title} {Nonintrusive nitric oxide density measurements in the effluent of core-heated airstreams},}\ }\href@noop {} {\bibfield  {journal} {\bibinfo  {journal} {AIAA Paper 90-1478}\ } (\bibinfo {year} {June 1990})}\BibitemShut {NoStop}%
\bibitem [{\citenamefont {Sharma}\ \emph {et~al.}(1992)\citenamefont {Sharma}, \citenamefont {Ruffin}, \citenamefont {Gillespie},\ and\ \citenamefont {Meyer}}]{Tau_Expanding5}%
  \BibitemOpen
  \bibfield  {author} {\bibinfo {author} {\bibfnamefont {S.}~\bibnamefont {Sharma}}, \bibinfo {author} {\bibfnamefont {S.}~\bibnamefont {Ruffin}}, \bibinfo {author} {\bibfnamefont {W.}~\bibnamefont {Gillespie}}, \ and\ \bibinfo {author} {\bibfnamefont {S.}~\bibnamefont {Meyer}},\ }\bibfield  {title} {\enquote {\bibinfo {title} {Nonequilibrium vibrational population measurements in an expanding flow using spontaneous raman spectroscopy},}\ }\href@noop {} {\bibfield  {journal} {\bibinfo  {journal} {AIAA Paper 92-2855}\ } (\bibinfo {year} {July 1992})}\BibitemShut {NoStop}%
\bibitem [{\citenamefont {Gillespie}\ \emph {et~al.}(1993)\citenamefont {Gillespie}, \citenamefont {Bershader}, \citenamefont {Sharma},\ and\ \citenamefont {Ruffin}}]{Tau_Expanding6}%
  \BibitemOpen
  \bibfield  {author} {\bibinfo {author} {\bibfnamefont {W.}~\bibnamefont {Gillespie}}, \bibinfo {author} {\bibfnamefont {D.}~\bibnamefont {Bershader}}, \bibinfo {author} {\bibfnamefont {S.}~\bibnamefont {Sharma}}, \ and\ \bibinfo {author} {\bibfnamefont {S.}~\bibnamefont {Ruffin}},\ }\bibfield  {title} {\enquote {\bibinfo {title} {Raman scattering measurements of vibrational and rotational distributions in expanding nitrogen},}\ }\href@noop {} {\bibfield  {journal} {\bibinfo  {journal} {AIAA Paper 93-0274}\ } (\bibinfo {year} {Jan 1993})}\BibitemShut {NoStop}%
\bibitem [{\citenamefont {Millikan}\ and\ \citenamefont {White}(1963)}]{Millikan-White}%
  \BibitemOpen
  \bibfield  {author} {\bibinfo {author} {\bibfnamefont {R.~C.}\ \bibnamefont {Millikan}}\ and\ \bibinfo {author} {\bibfnamefont {D.~R.}\ \bibnamefont {White}},\ }\bibfield  {title} {\enquote {\bibinfo {title} {Systematics of vibrational relaxation},}\ }\href@noop {} {\bibfield  {journal} {\bibinfo  {journal} {The Journal of Chemical Physics}\ } (\bibinfo {year} {1963})}\BibitemShut {NoStop}%
\bibitem [{\citenamefont {Grover}\ \emph {et~al.}(2019)\citenamefont {Grover}, \citenamefont {Schwartzentruber}, \citenamefont {Varga},\ and\ \citenamefont {Truhlar}}]{Grover_tau}%
  \BibitemOpen
  \bibfield  {author} {\bibinfo {author} {\bibfnamefont {M.~S.}\ \bibnamefont {Grover}}, \bibinfo {author} {\bibfnamefont {T.~E.}\ \bibnamefont {Schwartzentruber}}, \bibinfo {author} {\bibfnamefont {Z.}~\bibnamefont {Varga}}, \ and\ \bibinfo {author} {\bibfnamefont {D.~G.}\ \bibnamefont {Truhlar}},\ }\bibfield  {title} {\enquote {\bibinfo {title} {Vibrational energy transfer and collision-induced dissociation in $\text{O}+\text{O}_2$ collisions},}\ }\href@noop {} {\bibfield  {journal} {\bibinfo  {journal} {Journal of Thermophysics and Heat Transfer}\ } (\bibinfo {year} {2019})}\BibitemShut {NoStop}%
\end{thebibliography}%

\end{document}